\documentclass[journal,twoside,web]{ieeecolor}
\usepackage{tmi}

\let\labelindent\relax 

\usepackage{amssymb}
\usepackage{amsmath}
\usepackage{graphicx}
\usepackage[colorinlistoftodos]{todonotes}
\usepackage[colorlinks=true, allcolors=blue]{hyperref}
\usepackage{enumitem}

\usepackage{tikz}
\usepackage{tikzscale}
\usetikzlibrary{positioning,shapes,shapes.multipart}
\usetikzlibrary{decorations.pathreplacing}
\usetikzlibrary{arrows.meta}
\usetikzlibrary{calc}
\usetikzlibrary{patterns}
\usetikzlibrary{patterns.meta}

\usepackage{pgfplots}
\pgfplotsset{compat=1.5.1}
\usepgfplotslibrary{units}

\usepackage{bbm}
\newcommand{\IN}{\mathbbm{N}}		
\newcommand{\IR}{\mathbbm{R}}		

\renewcommand{\epsilon}{\varepsilon}				
\renewcommand{\theta}{\vartheta}                  	
\renewcommand{\phi}{\varphi}						

\newcommand{\set}[1]{\left\{ #1 \right\}}				
\newcommand{\floor}[1]{\left\lfloor #1 \right\rfloor}	
\newcommand{\ceil}[1]{\left\lceil #1 \right\rceil}		

\newcommand{\cmod}[1]{~\left(\textup{mod}\; #1 \right)}   

\newcommand{\signal}{\textup{Signal}}
\newcommand{\shift}{\textup{shift}}
\newcommand{\frameN}{\textup{frame}}

\newcommand{\mot}{\textup{mot}}
\newcommand{\virt}{\textup{virt}}

\newcommand{\Hann}{\textup{Hann}}
\newcommand{\meas}{\textup{meas}}

\newcommand{\zb}[1]{\mbox{\boldmath{${#1}$}}}

\renewcommand{\d}{\, \text{d}}

\usepackage[binary-units = true,exponent-product = \cdot]{siunitx}
\DeclareSIUnit{\mup}{\text{$\mu_0$}}

\definecolor{ibilight}{RGB}{193,216,237}
\definecolor{ibidark}{RGB}{0,73,146}
\definecolor{uke2}{RGB}{170,156,143}
\definecolor{uke3}{RGB}{87,87,86}
\definecolor{ukesec1}{RGB}{255,223,0}
\definecolor{ukesec2}{RGB}{239,123,5}
\definecolor{ukesec3}{RGB}{104,195,205}
\definecolor{ukesec4}{RGB}{138,189,36}
\definecolor{tuhh}{RGB}{45,198,214}

\author{Nadine Gdaniec, Marija Boberg, Martin M\"oddel, Patryk Szwargulski, Tobias Knopp
\thanks{This work was supported in part by the German Research Foundation (DFG) under Grant KN 1108/2-1 and in part by the Federal Ministry of Education and Research (BMBF) under Grant 05M16GKA and Grant 13XP5060B.}
\thanks{All authors are with the Section for Biomedical Imaging, University Medical Center Hamburg-Eppendorf, 20246 Hamburg, Germany and the Institute for Biomedical Imaging, Hamburg University of Technology, 21073 Hamburg, Germany (e-mail: t.knopp@uke.de).}
\thanks{Copyright (c) 2019 IEEE. Personal use of this material is permitted. However, permission to use this material for any other purposes must be obtained from the IEEE by sending a request to pubs-permissions@ieee.org.}}

\title{Suppression of Motion Artifacts Caused by Temporally
Recurring Tracer Distributions in Multi-Patch Magnetic Particle Imaging}

\begin{document}
\fontsize{9.5}{11.5}\selectfont

\maketitle

\begin{abstract}
    Magnetic particle imaging is a tracer based imaging technique to determine the spatial distribution of superparamagnetic iron oxide nanoparticles with a high spatial and temporal resolution. Due to physiological constraints, the imaging volume is restricted in size and larger volumes are covered by shifting object and imaging volume relative to each other. This results in reduced temporal resolution, which can lead to motion artifacts when imaging dynamic tracer distributions. A common source of such dynamic distributions are cardiac and respiratory motion in \textit{in-vivo} experiments, which are in good approximation periodic. We present a raw data processing technique that combines data snippets into virtual frames corresponding to a specific state of the dynamic motion. The technique is evaluated on the basis of measurement data obtained from a rotational phantom at two different rotational frequencies. These frequencies are determined from the raw data without reconstruction and without an additional navigator signal. The reconstructed images give reasonable representations of the rotational phantom frozen in several different states of motion while motion artifacts are suppressed.  
\end{abstract}


\section{Introduction}
    \IEEEPARstart{M}{otion} artifacts are a common patient-based problem in medical imaging caused by voluntary or involuntary patient movement during image acquisition. While voluntary movement can be reduced by immobilization or sedation, involuntary motion, such as cardiac or respiratory motion require technological solutions for the affected imaging systems. Even though numerous techniques have been developed to mitigate motion artifacts in magnetic resonance imaging~\cite{zaitsev2015} and computed tomography~\cite{boas2012} over the last decades research on this issue is still in its infancy for magnetic particle imaging (MPI).

    MPI is a tracer based imaging modality to determine the spatial distribution of superparamagnetic iron oxide nanoparticles with the help of static and dynamic magnetic fields~\cite{Gleich2005Nature}. MPI has proven to be suitable for medical applications like the detection of stroke~\cite{ludewig2017magnetic}, gut bleeding~\cite{yu2017magneticB}, cancer~\cite{arami2017tomographic}, stenosis~\cite{vaalma2017magnetic}, and the presence of cerebral aneurysms~\cite{sedlacik2016magnetic}. It was utilised for visualizing lung perfusion~\cite{zhou2017first}, labeled stem cells~\cite{bulte2015quantitative}, functional processes~\cite{cooley2018rodent}, and vascular interventions~\cite{salamon2016magnetic,haegele2016multi,herz2018magnetic}.

\begin{figure}
        \centering
        \input{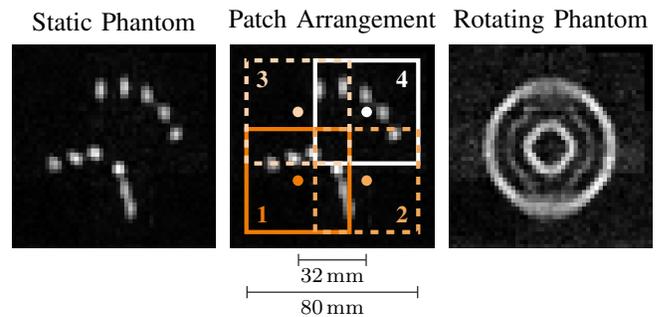}
        \caption{As reference, state-of-the-art reconstructions of a static (left) and a fast rotating phantom (right) are shown. The patch arrangement of the underlying multi-patch imaging sequence with corresponding patch numbers is depicted in the center. In the static case, the two diagonals consisting of three samples each are visible, as well as the five samples on the outer circular arc. In the dynamic case, however, the original point-shaped samples are distorted to circles and the structure of the phantom is completely lost, which illustrates the effects of motion artifacts in this image scenario.}
        \label{fig:stateoftheart}
    \end{figure}

    MPI has the potential to be a fast imaging technique with a repetition time in the millisecond range, which is fast enough for blood flow measurements~\cite{kaul2018magnetic}. In practice this only holds true for single imaging volumes. Peripheral nerve stimulation and tissue heating caused by the dynamic magnetic fields limit the field of view (FoV) of these volumes to only few centimeters for a gradient strength of the selection field larger than \SI{1}{\tesla \per \meter}~\cite{schmale2015mpi}, which is required for reaching spatial resolutions in the millimeter range. Techniques to increase the imaging volume either use a spatial shift of the fields~\cite{Knopp2015PhysMedBiol} or a mechanical shift of the object under examination~\cite{Goodwill2011TMI,szwargulski2018jmi} and are known as multi-patch sequences. Relative spatial shifts are performed with a much lower frequency compared to the excitation frequency to avoid heating and stimulation. These techniques have in common that they scan different areas of the object sequentially, which can reduce the temporal resolution to the point where motion artifacts can no longer be neglected, as illustrated in Fig.~\ref{fig:stateoftheart}.
    
    Currently, there are only a hand full of studies on the reduction of motion artifacts in MPI. In~\cite{gdaniec2017detection} a raw data processing technique has been proposed to detect motion and reduce the induced artifacts for single-patch imaging of temporally periodic tracer distributions. This technique has been extended to moving table multi-patch scenarios where the shifts are performed mechanically~\cite{szwargulski2018jmi}. Moreover, there are attempts to reduce motion artifacts using poly-rigid registration~\cite{ehrhardt2019}, which show promising results for single time points in a motion series. In this manuscript we generalize and enhance the technique proposed in~\cite{gdaniec2017detection} such that it can be used to reduce motion artifacts caused by temporally recurring tracer distributions for any MPI imaging sequence, i.e. single-patch and multi-patch. This makes it possible for the first time to reconstruct temporally recurring tracer distributions that are too fast to be resolved naively using multi-patch MPI.

\section{Problem Statement} \label{sec:problem}
    Currently, all reconstruction methods in MPI make the assumption that the measured signal is caused by a static particle distribution and a violation of this assumption can lead to motion artifacts in the reconstructed images. The strength of these artifacts is primarily determined by the speed at which the tracer distribution changes and the frame rate of the MPI measurement sequence. The motion artifacts we observed can be grouped into three categories: ghosting artifacts, smearing artifacts and patch boundary discontinuities. 
    
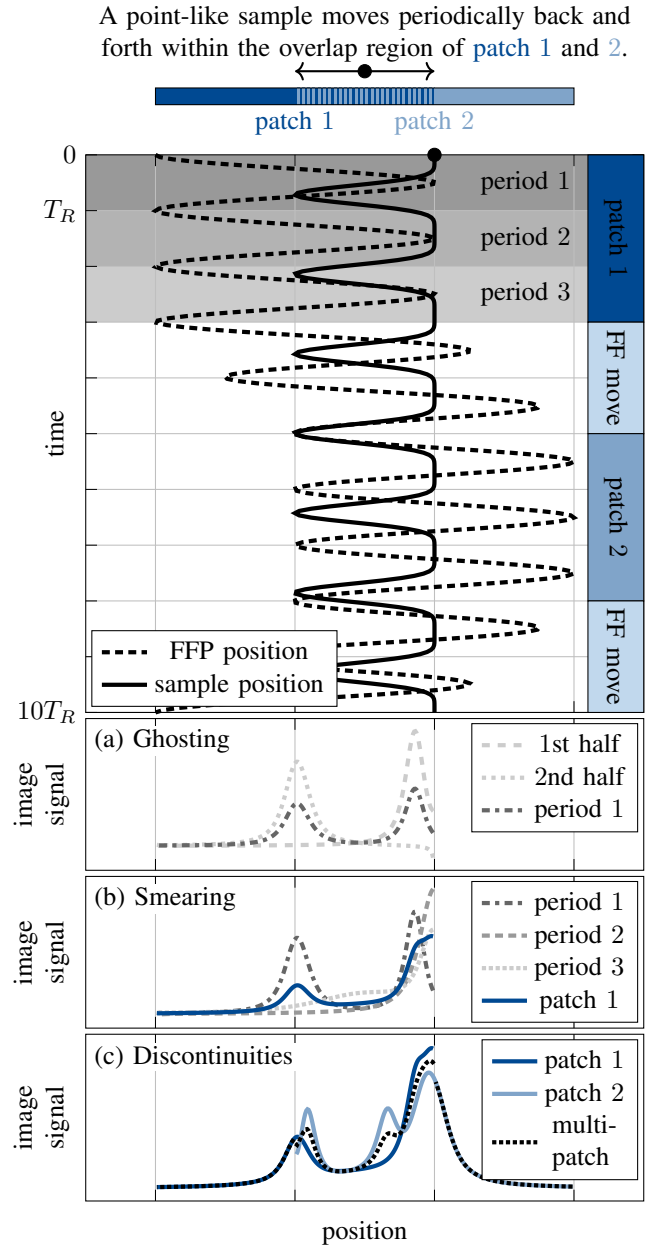
\begin{figure}
        \centering
        \begin{tikzpicture}[]
\pgfplotsset{xmin = {-20}, xmax = {20},
             xtick={-15,-5,5,15}, xticklabels={},
             grid=both,
             tick style={black},
             ylabel style={at={(-0.04,0.5)},anchor=base},
             }
\tikzdeclarepattern{name=mylines,
                    parameters={\pgfkeysvalueof{/pgf/pattern keys/size},
                                \pgfkeysvalueof{/pgf/pattern keys/angle},
                                \pgfkeysvalueof{/pgf/pattern keys/line width},},
                    bounding box={(0,-0.5*\pgfkeysvalueof{/pgf/pattern keys/line width}) and (\pgfkeysvalueof{/pgf/pattern keys/size}, 0.5*\pgfkeysvalueof{/pgf/pattern keys/line width})},
                    tile size={(\pgfkeysvalueof{/pgf/pattern keys/size},\pgfkeysvalueof{/pgf/pattern keys/size})},
                    tile transformation={rotate=\pgfkeysvalueof{/pgf/pattern keys/angle}},
                    defaults={size/.initial=5pt,
                                angle/.initial=45,
                                line width/.initial=.4pt,},
                    code={\draw[line width=\pgfkeysvalueof{/pgf/pattern keys/line width}](0,0) --  (\pgfkeysvalueof{/pgf/pattern keys/size},0);
                    },
                    }

\begin{axis}
    [
    clip=false,
    ylabel = {time},
    ymin = {0}, ymax = {10},
    width=9cm, height=9cm,
    ytick={0,1,2,3,4,5,6,7,8,9,10},yticklabels={$0$,$T_R$,,,,,,,,,$10T_R$}, 
    x dir=reverse, y dir=reverse,
    legend style={at={(0.01, 0.01)},anchor = south west},
    set layers]
    \addplot[fill=black!40,draw=none,forget plot,on layer=axis background] coordinates 
    {(20,0) (20,1) (-20,1) (-20,0)}; 
    \addplot[fill=black!30,draw=none,forget plot,on layer=axis background] coordinates 
    {(20,1) (20,2) (-20,2) (-20,1)}; 
    \addplot[fill=black!20,draw=none,forget plot,on layer=axis background] coordinates 
    {(20,2) (20,3) (-20,3) (-20,2)}; 
    
    \addplot+ [mark = {}, black, densely dashed, ultra thick] table [col sep=comma, x index = 1, y index = 0] {data/gdani2a.csv};
    \addlegendentry{FFP position}
    \addplot+ [mark = {}, black, ultra thick] table [col sep=comma, x index = 2, y index = 0] {data/gdani2a.csv};
    \addlegendentry{sample position}
    
    \draw[fill=ibidark] (axis cs:-16.0,0) rectangle (axis cs:-20.0,3) node[pos=.5,rotate=270] {patch $1$};
    \draw[fill=ibilight] (axis cs:-16.0,3) rectangle (axis cs:-20.0,5) node[pos=.5,rotate=270] {FF move};
    \draw[fill=ibidark!50] (axis cs:-16.0,5) rectangle (axis cs:-20.0,8) node[pos=.5,rotate=270] {patch $2$};
    \draw[fill=ibilight] (axis cs:-16.0,8) rectangle (axis cs:-20.0,10) node[pos=.5,rotate=270] {FF move};
    
    \node[fill=none, draw=none, inner sep=1pt, anchor=east] at (axis cs: -15,0.5) {period $1$}; 
    \node[fill=none, draw=none, inner sep=1pt, anchor=east] at (axis cs: -15,1.5) {period $2$};
    \node[fill=none, draw=none, inner sep=1pt, anchor=east] at (axis cs: -15,2.5) {period $3$};
    
    \draw[fill=ibidark!50, draw=none] (axis cs:-15.0,-0.9) rectangle (axis cs:5.0,-1.2) node[pos=.5,yshift=-0.35cm] {\textcolor{ibidark!50}{patch $2$}};
    \draw[fill=ibidark, draw=none] (axis cs:-5.0,-0.9) rectangle (axis cs:15.0,-1.2) node[pos=.5,yshift=-0.35cm] {\textcolor{ibidark}{patch $1$}};
    \draw[draw=none, pattern={mylines[size= 2pt,line width=1pt,angle=90]}, pattern color=ibidark!50] (axis cs: -5,-0.9) rectangle (axis cs: 5,-1.2);
    \draw[fill=none] (axis cs: -15,-0.9) rectangle (axis cs: 15,-1.2);
    \draw[fill=black] (axis cs: 0,-1.5) circle (2.5pt) node[above, text width=8cm, align=center] {A point-like sample moves periodically back and forth within the overlap region of \textcolor{ibidark}{patch $1$} and \textcolor{ibidark!50}{$2$}.};
    \draw[fill=black] (axis cs: -5,0) circle (2.5pt);
    \draw[<->,thick] (axis cs: -5,-1.5) -- (axis cs: 5,-1.5);
\end{axis}

\begin{axis}
    [yshift=-2.1cm,
    x dir=reverse,
    ylabel style={align=center},
    ylabel = {image\\signal}, 
    ytick=\empty, 
    width=9cm, height=3.6cm, 
    legend style={at={(0.99,0.97)},anchor=north east}]
    \node[anchor=north west] at (rel axis cs: 0.01,0.95) [draw=none, fill=white, inner sep=1pt] {(a) Ghosting};
    \addplot+ [mark = {}, black!20, ultra thick, dashed] table [col sep=comma, x index = 0, y index = 1] {data/gdani2b.csv};
    \addlegendentry{$1$st half}
    \addplot+ [mark = {}, black!20, ultra thick, dotted] table [col sep=comma, x index = 0, y index = 2] {data/gdani2b.csv};
    \addlegendentry{$2$nd half}
    \addplot+ [mark = {}, black!60, ultra thick, dashdotted] table [col sep=comma, x index = 0, y index = 3] {data/gdani2b.csv};
    \addlegendentry{period $1$}
\end{axis}

\begin{axis}
    [yshift=-4.2cm,
    x dir=reverse,
    ylabel style={align=center},
    ylabel = {image\\signal},
    ytick=\empty, 
    width=9cm, height=3.6cm,
    legend style={at={(0.99, 0.97)},anchor = north east}]
    \node[anchor=north west] at (rel axis cs: 0.01,0.95) [draw=none, fill=white, inner sep=1pt] {(b) Smearing};
    \addplot+ [mark = {}, black!60, ultra thick, dashdotted] table [col sep=comma, x index = 0, y index = 1] {data/gdani2c.csv};
    \addlegendentry{period $1$}
    \addplot+ [mark = {}, black!40, ultra thick, densely dashed] table [col sep=comma, x index = 0, y index = 2] {data/gdani2c.csv};
    \addlegendentry{period $2$}
    \addplot+ [mark = {}, black!20, ultra thick, densely dotted] table [col sep=comma, x index = 0, y index = 3] {data/gdani2c.csv};
    \addlegendentry{period $3$}
    \addplot+ [mark = {}, ibidark, ultra thick] table [col sep=comma, x index = 0, y index = 4] {data/gdani2c.csv};
    \addlegendentry{patch $1$}
\end{axis}

\begin{axis}
    [yshift=-6.5cm,
    x dir=reverse,
    xlabel = {position}, 
    ylabel style={align=center},
    ylabel = {image\\signal},
    xlabel style={yshift=0.1cm},
    ytick=\empty,  
    width=9cm, height=3.8cm,
    legend style={at={(0.99,0.97)}, anchor=north east},cells={align=left},]
    \node[anchor=north west] at (rel axis cs: 0.01,0.95) [draw=none, fill=white, inner sep=1pt] {(c) Discontinuities};
    \addplot+ [mark = {}, ibidark, ultra thick] table [col sep=comma, x index = 0, y index = 1] {data/gdani2d.csv};
    \addlegendentry{patch $1$}
    \addplot+ [mark = {}, ibidark!50, ultra thick] table [col sep=comma, x index = 2, y index = 3] {data/gdani2d.csv};
    \addlegendentry{patch $2$}
    \addplot+ [restrict x to domain=-15:-5, mark={}, black, ultra thick, densely dotted] table [col sep=comma, x index = 4, y index = 5] {data/gdani2d.csv};
    \addlegendentry{multi-\\patch}
    \addplot+ [restrict x to domain=-5:5, mark={}, black, ultra thick, densely dotted] table [col sep=comma, x index = 4, y index = 5] {data/gdani2d.csv};
    \addplot+ [restrict x to domain=5:15, mark={}, black, ultra thick, densely dotted] table [col sep=comma, x index = 4, y index = 5] {data/gdani2d.csv};
\end{axis}
\end{tikzpicture}
        \vspace{-0.5cm}
        \caption{At the top most plot the FFP movement of an exemplary 1D multi-patch sequence and the movement of a point like sample are shown. The basic repetition time of the FFP movement is $T_R$. Each of the two patches is scanned for three cycles (patch $1$ and patch $2$) before the focus fields change focus within the next two cycles (FF move). Reconstructions of the first and second half of the signal yield different images as shown in plot (a) for the first excitation period. Both location as well as strength of the image signal differ due to the fast motion of the sample. A standard reconstruction of the average measurement signal of the first period will therefore result in a ghosting artifact (period $1$), where the image shows two samples instead of one. Separate reconstructions of the three periods of patch $1$ (light, medium, and dark gray) are shown in plot (b). Here too, the object movement causes the sample to be reconstructed in different locations. These differences accumulate in ghosting and smearing artifacts in the reconstruction of the average signal of the whole patch (patch $1$). Separate reconstructions of the first two patches are shown in plot (c). Apart from the motion artifacts mentioned in the previous cases, one observes different signal intensities where the image domains overlap. These lead to discontinuities at the patch boundaries in the reconstructed image of a complete multi-patch cycle in case the discontinuities are not handled separately. While the other motion artifacts can occur in single-patch imaging, the latter is specific to multi-patch MPI.}
        \label{fig:problem}
    \end{figure}

    The following example should serve as an illustration of the different types of artifacts. To this end, consider a 1D multi-patch sequence with two patches as illustrated in Fig.~\ref{fig:problem}. The basic cosine excitation is done by a field free point (FFP) moving back and forth with a repetition time of $T_R$. Each patch is scanned for three cycles before the focus fields change focus within the next two cycles resulting in a total repetition time of $10T_R$ for the complete multi-patch sequence. Moreover, we consider a point-like sample moving periodically back and forth within the overlap region of the patches.
    
    \begin{enumerate}[label=(\alph*)]
        \item \textit{Ghosting artifacts:} Whenever the FFP traverses the sample a strong MPI signal is generated encoding the location of the sample. During one excitation period the FFP traverses each position twice. In our example, FFP-sample crossings appear at different positions due to the fast motion of the sample. This can be seen when reconstructing the first and the second half of the first excitation period independently. The net effect on the reconstruction of the average signal of the full period is a ghosting artifact, where the sample appears twice in the reconstructed image as shown in Fig.~\ref{fig:problem}~(a).
        \item \textit{Smearing artifacts:} If at least two of these ghosting artifacts are overlapping they blend into a single broader signal, which we refer to as smearing artifact. These artifacts can be especially prominent when multiple FFP-sample transits occur only slightly displaced. For our fast moving sample this can be observed in on the right hand side of Fig.~\ref{fig:problem}~(b).
        \item \textit{Patch boundary discontinuities:} Considering the reconstructed images of different patches, each may yield different signals in the region of patch overlap, which can be observed for our example. Therefore, a multi-patch reconstruction of the complete measurement without boundary continuation handling can cause discontinuities at the patch boundary if the signals from different patches differ at these locations as shown in Fig.~\ref{fig:problem}~(c).
    \end{enumerate}

\section{Theory}
    The method for reducing motion-induced artifacts presented in this manuscript improves and generalizes the method for single-patch measurements presented in~\cite{gdaniec2017detection}. To recapitulate the latter, consider a periodically changing tracer distribution, discretize one motion period into a finite number of motion states, and group the data snippets by these states. The reoccurring data snippets of an arbitrary state of motion do not necessarily cover a complete excitation cycle, which is necessary for reconstruction. This issue is resolved by processing multiple raw data snippets into a single virtual frame, which represents a single-patch measurement, where the distribution is frozen in the considered state.
    
    We generalize and improve the aforementioned method in three different ways.
    \begin{itemize}
        \item Depending on whether the length of the data snippets is shorter or longer than the repetition time of the excitation, two different processing approaches were used in~\cite{gdaniec2017detection}. We combine both in a unified framework.
        \item Changes in the tracer distribution within a sampling interval lead to discontinuities within the virtual frame causing image artifacts. We propose to reduce these by a spectral leakage correction.
        \item We generalize the method to the multi-patch case. I.e. we describe how to process data snippets of a specific state of motion acquired during a multi-patch measurement into a virtual multi-patch frame.
    \end{itemize}
    After sketching our approach we describe the procedure in more detail using mathematical notation.

    \subsection{MPI Signal: Static Object} \label{static}
        First, we focus on the MPI signal of a static object fitting into a single patch. The relation between the static distribution $c_{s}:\IR^3 \rightarrow \IR_+$, which is the image to be reconstructed, and the MPI signal $u:[0,LT_R]\rightarrow \IR$ can be described by
        \begin{equation}
            u(t) = -\mu_0 \int\limits_{\IR^3}\zb{p}(\zb{r})\cdot\frac{\partial{\zb{m}}}{\partial{t}}(\zb{r},t)c_{s}(\zb{r})\d^3 r.
            \label{eq:signalequation}
        \end{equation}
        This equation contains the vacuum permeability $\mu_0$, the coil sensitivity of the receive coil $\zb p:\IR^3 \rightarrow \IR^3$, the magnetization of the tracer $\zb m:\IR^3 \times \IR\rightarrow \IR^3$, the cycle duration of the drive-field excitation $T_R$, and the number of measured cycles $L$. Due to the periodicity of the excitation field 
        \begin{equation}\label{eq:excitationPeriodicity}
            \frac{\partial{\zb{m}}}{\partial{t}}(\zb{r},t+qT_R)=\frac{\partial{\zb{m}}}{\partial{t}}(\zb{r},t),
        \end{equation} 
        we have a $T_R$-periodic measurement signal, i.e.
        \begin{equation*}
            u(t+qT_R)=u(t),
        \end{equation*}
        for all $t\in [0,T_R]$ and $q \in \{0,1,\dots,L-1\}$. The discrete time signal acquired during one drive-field cycle can be used to reconstruct an image of the static object.

    \subsection{MPI Signal: Object with Movement} \label{movement}
        During \textit{in-vivo} measurements, the object is not static, but experiences motion. The motion is caused e.g.~by respiration motion, cardiac motion, or the motion of MPI-tracer-coated interventional devices. The repetition time $T_R$ in MPI is quite small (e.g.~$\SI{21.54}{\milli \second}$ for a single imaging volume) resulting in a high temporal resolution for single-patch imaging without averaging. Quasi-static tracer distributions are those with only slight displacements during $T_R$ and can be handled like static data. In other case (dynamic tracer distributions), this data handling might introduce severe motion artifacts into the processed signal and therefore into the final MPI tomograms as discussed in section~\ref{sec:problem}.

        If we assume a continuously differentiable dynamic tracer distribution $c(\zb{r},t)$ over time, the generated signal is given by
        \begin{align*}
            u(t) &= -\mu_0\negthickspace\int\limits_{\IR^3}\negmedspace\zb{p}(\zb{r})\cdot
            \frac{\partial{}}{\partial{t}}\Big(\zb{m}(\zb{r},t)c(\zb{r},t)\Big)\d^3 r\\
            &= -\mu_0\negthickspace\int\limits_{\IR^3}\negmedspace\zb{p}(\zb{r})\cdot
            \negmedspace\left(\negmedspace\frac{\partial{\zb m}}{\partial{t}}(\zb{r},t)c(\zb{r},t)+\zb m(\zb{r},t)\frac{\partial{c}}{\partial{t}}(\zb{r},t)\negmedspace\right)\negthinspace\d^3 r,
        \end{align*}
        and is in general no longer $T_R$-periodic. In this work we limit the considered dynamic tracer distributions to those whose temporal change provides negligible contributions to the measurement signal, i.e.
        \begin{equation}
            u(t) \approx -\mu_0\int\limits_{\IR^3}\zb{p}(\zb{r})\cdot \frac{\partial{\zb m}}{\partial{t}}(\zb{r},t)c(\zb{r},t)\d^3 r.
            \label{eq:slowdynamic}
        \end{equation}
        In most cases this will only be a mathematical. In practice there are only very few MPI systems, where excitation is done by mechanical motion of the sample~\cite{vogel2015shot}.
        
        In order to discretize the dynamics of the tracer distribution we define $\Delta t > 0$ to be the largest duration for which the tracer distribution remains approximately static, i.e.
        \begin{equation}
            c(\zb{r},t) \approx c(\zb{r},t_s),\quad t\in I(t_s,\Delta t),
            \label{eq:approxStaticDistr}
        \end{equation}
        where $I(t,\Delta t) := [t,t+\Delta t]$ is a short notation for a length $\Delta t$ time interval starting from $t$. The time point $t_s$ is associated with a certain state of the motion. In that case the signal snippet generated by the dynamic distribution $c(\zb{r},t)$ during $I(t_s,\Delta t)$, is approximately that of a static distribution $c(\zb{r},t_s)$
        \begin{align}
            \begin{split}
                u(t) &\overset{\footnotesize \eqref{eq:slowdynamic}}{\approx} -\mu_0 \!\int\limits_{\IR^3}\!\zb{p}(\zb{r})\cdot\frac{\partial{\zb{m}}}{\partial{t}}(\zb{r},t)c(\zb{r},t)\d^3 r \\
                &\overset{\footnotesize \eqref{eq:approxStaticDistr}}{\approx} -\mu_0 \!\int\limits_{\IR^3}\!\zb{p}(\zb{r})\cdot\frac{\partial{\zb{m}}}{\partial{t}}(\zb{r},t)c(\zb{r},t_s)\d^3 r \\
                &= u_{t_s}^{\Delta t}(t), \qquad \qquad \qquad \qquad \qquad t \in I(t_s,\Delta t).
            \end{split}\label{eq:quasiStatic}
        \end{align}
        
    \subsection{Data Grouping} \label{sec:datasnippets}
        To be able to reconstruct the tracer distribution $c(\zb{r},t_s)$ at time $t_s$, a complete drive-field cycle describing the static tracer distribution is required. This is fulfilled for the quasi-static approximation ($\Delta t > T_R$) but requires further investigation in the other cases.
        
        In this work, we restrict to temporally recurring tracer distributions which are dynamic tracer distributions satisfying
        \begin{align}
            \begin{split}
                c(\zb{r},\rho(t,n))&\approx c(\zb{r},t), \quad t \in[0,T_\mot],  \; n\in\IN_0.
            \end{split}
            \label{eq:repeatDistr}
        \end{align}
        The mapping $\rho:\IR\times \IN_0 \rightarrow \IR$ provides the temporal distance between $0$ and the $n$-th repetition of the motion state at $t_s$ and $T_\mot$ is the duration of the first motion cycle. In particular, we have $\rho(t_s,0)=t_s$ and $\rho(t_s,1) = t_s+T_\mot$. In practice, the parameter $n$ is limited by the maximum number of repetitions $N_{t_s} = \max\set{n\in\IN_0 : \rho(t_s,n) \leq LT_R }$ that occur during the acquisition of the MPI signal. A special case is a $T_\mot$-periodic dynamic, where $\rho(t_s,n)=t_s+n T_\mot$ and without loss of generality $t_s\in [0,T_\mot]$. The more general definition \eqref{eq:repeatDistr} includes realistic scenarios, like those we have \textit{in-vivo}, where the heart beat and respiration are slightly irregular in which case a single period length $T_\mot$ does not describe the system accurately (especially if we measure for longer durations). Slight variations of the motion frequency result in different time intervals that fulfill the quasi-static approximation. The highest frequency results in most prominent artifacts and thus the time interval $\Delta t$ is chosen relative to the highest occurring motion frequency.
        
        Using equation~\eqref{eq:repeatDistr} we collect all data snippets where the dynamic distribution $c(\zb{r},t)$ is approximately equal to the motion state $c(\zb r,t_s)$. For $t \in I(\rho(t_s,n),\Delta t)$ we can approximate $u(t)$ by 
        \begin{align*}
            \begin{split}
                u(t) &\overset{\footnotesize \eqref{eq:slowdynamic}}{\approx} -\mu_0 \!\int\limits_{\IR^3}\!\zb{p}(\zb{r})\cdot\frac{\partial{\zb{m}}}{\partial{t}}(\zb{r},t)c(\zb{r},t)\d^3 r \\
                &\overset{\footnotesize \eqref{eq:approxStaticDistr}}{\approx} -\mu_0 \!\int\limits_{\IR^3}\!\zb{p}(\zb{r})\cdot\frac{\partial{\zb{m}}}{\partial{t}}(\zb{r},t)c(\zb{r},\rho(t_s,n))\d^3 r
                \\ &\overset{\footnotesize \eqref{eq:repeatDistr}}{\approx} -\mu_0 \!\int\limits_{\IR^3}\!\zb{p}(\zb{r})\cdot\frac{\partial{\zb{m}}}{\partial{t}}(\zb{r},t)c(\zb{r},t_s)\d^3 r \\
                &= u_{t_s}^{\Delta t}(t)
            \end{split}
        \end{align*}
        where $n \in \set{0,\dots,N_{t_s}}$ can be seen as a label for the $n$-th data snippet. For $t \in [0,LT_R]\setminus\bigcup_{n = 0}^{N_{t_s}} I(\rho(t_s,n),\Delta t)$ we set $u_{t_s}^{\Delta t}(t) = 0$.
        To this end, a masking function $\sigma^{\Delta t}_{t_s}$ is introduced that masks all aforementioned data snippets in the measured time signal $u(t)$. The function $\sigma^{\Delta t}_{t_s}$ can be defined as
        \begin{align}
          \begin{split}
            &\sigma^{\Delta t}_{t_s}:\IR \rightarrow \set{0,1},\\ 
            &t\mapsto 
            \begin{cases} 
            1, & t\in\bigcup\limits_{n=0}^{N_{t_s}} I(\rho(t_s,n),\Delta t)\\
            0, & \text{else} 
            \end{cases}.
          \end{split} \label{eq:sigma}
        \end{align}
        Thus, $u_{t_s}^{\Delta t}(t) \approx \sigma_{t_s}^{\Delta t}(t) u(t)$ for all $t\in [0,LT_R]$.
        
    \subsection{Data Combination}
        As pointed out in the last paragraph of this section the data snippets do not necessarily cover a complete excitation cycle in which case they can not be directly used for reconstruction. In addition, as the number of data snippets increases, there are redundancies in the data that can be exploited. Both issues can be resolved by combining the snippets into virtual frames~\cite{gdaniec2017detection}. One frame for each state of motion. 
        
        Using our collected data snippets, we can now define the virtual frame by
        \begin{equation}
	        \bar{u}^\text{virt}_{t_s}(t) = \frac{\sum\limits_{l=0}^{L-1}
	        u_{t_s}^{\Delta t}(t+lT_R)}
	        {\sum\limits_{l=0}^{L-1}\sigma^{\Delta t}_{t_s}(t+lT_R)},\quad t\in [0,T_R].
	    \label{eq:fast}
        \end{equation}
        Here, we need to divide the averaged signal by the number of averages that are taken for each time point $t$.
        
        One potential pitfall is that equation~\eqref{eq:fast} is undefined if
        \begin{equation}
            \exists \,t\in [0,T_R]: \sum\limits_{l=1}^{L-1}\sigma^{\Delta t}_{t_s}(t+l T_R)=0,
            \label{eq:condition}
        \end{equation}
        since the nominator and denominator would be zero for these $t$. This means that for the motion state at $t_s$ the measured data snippets do not suffice to fill the entire interval $[0,T_R]$ of the virtual frame. To fix this issue the measurement sequence needs to be extended by increasing the number of measured periods $L$.

    \subsection{Spectral Leakage Correction} \label{spectralLeakage}
        The approaches discussed so far assumed that the tracer distribution is static within the time interval $I(t_s,\Delta t)$. In practice this is only an approximation resulting in discontinuities at the edges of the pieces added to the virtual frame. Even small discontinuities can worsen the reconstruction quality drastically, since they lead to spectral leakage in the Fourier space, which is usually considered for reconstruction.
            
        To address this issue we propose to window the signal prior to filling the virtual frame. To this end, a window function $h:\IR\rightarrow [0,1]$ with support $\text{supp}(h) = [0,1]$ is defined, which can for instance be the Hann window
        \begin{equation*}
            h^\text{Hann}:\IR \rightarrow [0,1],\quad t\mapsto \begin{cases} \frac{1}{2}\left(1-\cos 2 \pi t\right), & t \in [0, 1]\\
            0, & \text{else}
        \end{cases}, 
        \end{equation*}
        or the rectangular window
        \begin{equation*}
            h^\text{Rect}:\IR \rightarrow [0,1],\quad t\mapsto \begin{cases} 1, & t \in [0, 1]\\
            0, & \text{else}
            \end{cases}.  
        \end{equation*}
        The specific choice of the window function has an influence on the efficiency of leakage correction~\cite{weber2016iwmpi,weberThesis}. 
        
\begin{figure}
            \centering
            \begin{tikzpicture}
\pgfmathsetmacro{\sizeX}{6}
\pgfmathsetmacro{\sizeY}{0.8}
\pgfmathsetmacro{\shiftY}{-1.3}
\tikzstyle{nodeU} = [inner sep = 1pt, fill=white]
\pgfmathsetmacro{\shiftU}{\sizeY / 2}

\tikzset{
  >={Latex[width=2mm,length=2mm]},}

\pgfplotsset{
    scale only axis,
 	view={0}{90},
 	ytick=\empty,
 	xtick={0,1}, 
    xticklabel = \empty,
    footnotesize,
    enable tick line clipping=false,
 	x tick label style = {yshift=0.05cm},
    xtick pos = bottom, xtick align = outside,
    colormap name = viridis,
    width=\sizeX cm,
    height=\sizeY cm,
    xmin=0,xmax=1,
    ylabel shift = 0.1cm,
	}

\draw[draw=none,fill=ibilight!80] (0,5.5*\shiftY) rectangle (1.2,\sizeY); 
\draw[draw=none,fill=ibilight!60] (5.13,5.5*\shiftY) rectangle (\sizeX,\sizeY+2*\shiftY); 
\draw[draw=none,fill=ibilight!40] (2.76,5.5*\shiftY) rectangle (3.96,\sizeY+3.5*\shiftY); 

	\begin{scope}
        \begin{axis}[
            xticklabels={$0$,$T_R$},
            ylabel=$u(t)$,
            extra x ticks = {0.2}, 
            extra x tick labels = {{$t_s+\Delta t$}},
            axis y line*=left,
            ]
            \addplot3[surf, shader=interp] 
                table [x index = 0, z index=2, y index = 1] {data/gdani3a.txt};
        \end{axis}
        
        \begin{axis}[ scale only axis, 
            axis y line*=right, 
            axis x line=none,
            ymin=-300,ymax=300]
            \addplot[white,thick] table [col sep=comma, x index = 0, y index = 6] {data/gdani3b.txt};
        \end{axis}
    \end{scope}
    
	\begin{scope}[yshift=\shiftY cm]
        \begin{axis}[
            xticklabels={$T_R$,$2T_R$},
            ylabel=$u(t)$,
            axis y line*=left,
            zmin=0, zmax=1,
            ]
            \addplot3[surf, shader=interp,] 
                table [x index = 0, z index=2, y index = 1] {data/gdani3c.txt};
        \end{axis}
    
        \begin{axis}[ scale only axis, 
            axis y line*=right, 
            axis x line=none,
            ymin=-300,ymax=300]
            \addplot[white,thick] table [col sep=comma, x index = 0, y index = 7] {data/gdani3b.txt};
        \end{axis}
    \end{scope}
    
	\begin{scope}[yshift=2*\shiftY cm]
        \begin{axis}[
            xticklabels={$2T_R$,$3T_R$},
            ylabel=$u(t)$,
            extra x ticks = {0.855,1.055}, 
            extra x tick labels = {{$\rho (t_s,1)$}, {$\rho (t_s+\Delta t,1)$}}, 
            axis y line*=left,
            ]
            \addplot3[surf, shader=interp,] 
                table [x index = 0, z index=2, y index = 1] {data/gdani3d.txt};
        \end{axis}
    
        \begin{axis}[ scale only axis, 
            axis y line*=right, 
            axis x line=none,
            ymin=-300,ymax=300]
            \addplot[white,thick] table [col sep=comma, x index = 0, y index = 8] {data/gdani3b.txt};
        \end{axis}
    \end{scope}
    
	\begin{scope}[yshift=3.5*\shiftY cm]
     	\node at (0.5*\sizeX,1.5) {$\vdots$};
    	
        \begin{axis}[
            xticklabels={$(L-1)T_R$,$LT_R$},
            ylabel=$u(t)$,
            extra x ticks = {0.46,0.66}, 
            extra x tick labels = {{\hspace*{-10pt}$\rho (t_s,N_{t_s})$}, {\hspace*{20pt}$\rho (t_s+\Delta t,N_{t_s})$}}, 
            axis y line*=left,
            ]
            \addplot3[surf, shader=interp,] 
                table [x index = 0, z index=2, y index = 1] {data/gdani3e.txt};
        \end{axis}
    
        \begin{axis}[ scale only axis, 
            axis y line*=right, 
            axis x line=none,
            ymin=-300,ymax=300]
            \addplot[white,thick] table [col sep=comma, x index = 0, y index = 9] {data/gdani3b.txt};
        \end{axis}
        
        \foreach \x in {0.6,5.73,3.36} {
        		\draw[->] (\x, -0.8) -- (\x,-1.4);
        };
    \end{scope}
    
	\begin{scope}[yshift=5.5*\shiftY cm]
    
        \begin{axis}[ scale only axis, 
            xticklabels={$0$,$T_R$},
            ylabel=$\bar{u}^\virt_{t_s,h^\Hann}(t)$,
            xlabel = $t$, x unit = s, unit markings=slash space,
            xlabel style = {yshift=0.3cm},
            ymin=-300,ymax=300]
            \addplot[black,thick] table [col sep=comma, x index = 0, y index = 1] {data/gdani3b.txt};
        \end{axis}
    \end{scope}

	\pgfmathsetmacro{\heightCb}{-3.5*\shiftY + \sizeY}
	\begin{scope}[yshift=\sizeY cm,xshift=\sizeX cm + 0.1 cm]
    	\begin{axis}[
    	    clip=true,
    	    hide axis,
    	    scale only axis,
    	    height=0pt, width=0pt,
    	    colormap name = viridis,
    	    colorbar,
    	    point meta min=0, point meta max=1,
    	    colorbar style={
    			x = 0.3cm,		
    			y = \heightCb cm,
    			ytick distance = 0.2,
    	        ylabel = {$\omega^{\Delta t}_{t_s,h^\textup{Hann}}(t)$},
    			ylabel style={rotate=180, at={(2.7,0.5)},anchor=south},
    	    },
    	    ]
    	    \addplot [draw=none] coordinates {  (0,0)
    	                                        (1,1)   };
    	\end{axis}
	\end{scope}
\end{tikzpicture}
            \caption{A graphical representation of the combination of data snippets into a virtual frame is shown. The measured data (white) is subdivided with respect to the $L$ drive-field periods. Multiplication with the window function $\omega_{t_s,h^\Hann}^{\Delta t}$ selects the data snippets corresponding to the selected motion state at $t_s = 0$ and reduces spectral leakage. Summation of all windowed data snippets and subsequent normalization yields the virtual frame $\bar{u}_{t_s,h^\Hann}^\virt$ as indicated by blue rectangles for the data snippets shown.}
            \label{fig:virtualFrame}
        \end{figure}
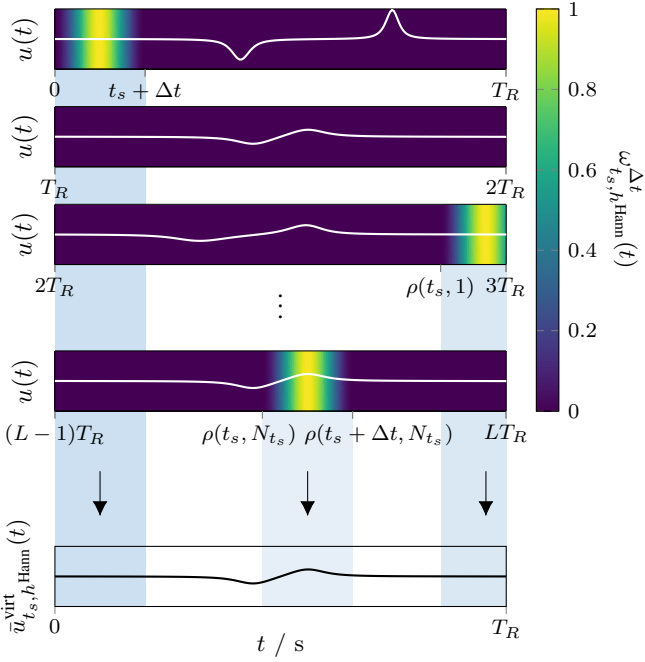

        We scale the window and define
        \begin{equation*}
            h_{\Delta t}:\IR\rightarrow [0,1],\quad t\mapsto  h{\left(\frac{t}{\Delta t}\right)},  
        \end{equation*}
        where the support of $h_{\Delta t}$ is  $\text{supp}(h_{\Delta t}) = [0, \Delta t]$. To apply the spectral leakage correction we replace the non-zero parts of the binary function $\sigma^{\Delta t}_{t_s}$ with a window that has exactly the width $\Delta t$. This modified function can be defined by using a sum of shifted window functions
        \begin{equation}
            \omega^{\Delta t}_{t_s, h}:\IR \rightarrow [0,1],\quad
            t\mapsto  \sum_{n=0}^{N_{t_s}} h_{\Delta t}(t - \rho(t_s,n) ).
            \label{eq:window}
        \end{equation}
        By inserting the rectangular window, one can see that $\omega^{\Delta t}_{t_s, h^\text{Rect}} = \sigma^{\Delta t}_{t_s}$, which implies that $\omega^{\Delta t}_{t_s, h}$ includes $\sigma^{\Delta t}_{t_s}$ as a special case when the rectangular window is chosen. We define the spectral leakage corrected virtual frame by
        \begin{equation}
	        \bar{u}_{t_s, h}^\virt(t) = \frac{\sum\limits_{l=0}^{L-1}u(t+lT_R)\omega^{\Delta t}_{t_s, h}(t+lT_R)}{\sum\limits_{l=0}^{L-1}\omega^{\Delta t}_{t_s, h}(t+lT_R)},\quad t\in [0,T_R]
	    \label{eq:averagedAndSLCorr}
        \end{equation}
        as illustrated in Fig.~\ref{fig:virtualFrame}. Again, equation~\eqref{eq:averagedAndSLCorr} is undefined if condition~\eqref{eq:condition} holds and  the number of repetitions $L$ needs to be increased in that case. Henceforth, we will refer to equation~\eqref{eq:averagedAndSLCorr} when using the notion virtual frame, since equation~\eqref{eq:averagedAndSLCorr} is a strict generalization of equation~\eqref{eq:fast}.

    \subsection{Multi-Patch Imaging Sequence} \label{Sec:multiPatch} \label{sec:MultiPatchMapping}
\begin{figure}
            \centering
            \begin{tikzpicture}[node distance=1pt,
                    every node/.style={fill=white, font=\scriptsize, inner sep=0pt}, align=center,decoration=brace]
\tikzset{
    base/.style = {rectangle, rounded corners, draw=black,
                 text width=1cm, minimum height=0.5cm,
                 text centered}, 
    patch1/.style = {base, fill=ibidark!100},
    patch2/.style = {base, fill=ibidark!85},
    patch3/.style = {base, fill=ibidark!70},
    patch4/.style = {base, fill=ibidark!55},
    move/.style = {base,fill=ibilight,minimum width=1cm,}
}

    \node (patch1)  [patch1]   {Patch 1};
    \node (pause1)  [move, right = of patch1]  {FF move};
    \node (patch2)  [patch2,right = of pause1]  {Patch 2};
    \node (pause2)  [move, right =  of patch2]  {FF move};
    \node (patch3)  [patch3,right = of pause2]  {Patch 3};
    \node (pause3)  [move, right = of patch3]  {FF move};
    \node (patch4)  [patch4,right = of pause3]  {Patch 4};
    \node (pause4)  [move, right = of patch4]  {FF move};
    \node (patch1_2)   [patch1, below = 0.8cm of patch1]   {Patch 1};
    \node (pause1_2)   [move,  right = of patch1_2]   {FF move};
    \node (patch2_2)   [patch2,right = of pause1_2]   {Patch 2};
    \node (pause2_2)   [move,  right = of patch2_2]   {FF move};
    \node (patch3_2)   [patch3,right = of pause2_2]   {Patch 3};
    \node (pause3_2)   [move,  right = of patch3_2]   {FF move};
    \node (patch4_2)   [patch4,right = of pause3_2]   {Patch 4};
    \node (pause4_2)   [move,  right = of patch4_2]   {FF move};
    \node (patch1_3)   [patch1, below = 1.1cm of patch1_2] {Patch 1};
    \node (pause1_3)   [move,  right = of patch1_3]   {FF move};
    \node (patch2_3)   [patch2,right = of pause1_3]   {Patch 2};
    \node (pause2_3)   [move,  right = of patch2_3]   {FF move};
    \node (patch3_3)   [patch3,right = of pause2_3]   {Patch 3};
    \node (pause3_3)   [move,  right = of patch3_3]   {FF move};
    \node (patch4_3)   [patch4,right = of pause3_3]   {Patch 4};
    \node (pause4_3)   [move,  right = of patch4_3]   {FF move};
    
    \node()[below right = of pause2_2,xshift=-0.5pt,yshift=-0.15cm]{\vdots};
    
    \node () [left = 8pt of patch1, rotate=90, xshift=0.5cm] {Frame $1$};
    \node () [left = 8pt of patch1_2, rotate=90, xshift=0.5cm] {Frame $2$};
    \node () [left = 8pt of patch1_3, rotate=90, xshift=0.5cm] {Frame $F$};
    
    \draw[|-|]  ($(patch1)+(-0.5,-0.5)$) -- node[below=0.1cm] {$L_P T_R$}  ($(patch1)+(0.5,-0.5)$);
    \draw[|-|]  ($(pause1)+(-0.5,0.5)$) -- node[above=0.1cm] {$T_\shift$}  ($(pause1)+(0.5,0.5)$);
    \draw[|-|]  ($(patch1_3)+(-0.5,-0.5)$) -- node[below=0.1cm] {$T_\frameN$}  ($(pause4_3)+(0.5,-0.5)$);

\end{tikzpicture}
            \caption{Illustration of a multi-patch imaging sequence with four patches. At each patch position $L_{P}$ drive-field excitation cycles with duration $T_R$ will be run through. Following the focus fields change (FF move) in a fixed time $T_\text{shift}$. The total time of a single cycle multi-patch sequence is $T_\text{frame}$ and the data acquired during that time are referred to as one frame.}
            \label{fig:multiPatchSequence}
        \end{figure}
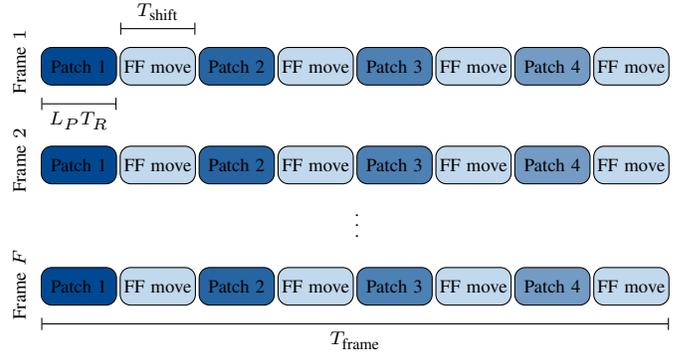

        To increase the FoV in MPI, multiple patches shifted relative to each other are measured sequentially. The number of patches is denoted by $P$. An exemplary temporal sequence of the measurement process is illustrated in Fig.~\ref{fig:multiPatchSequence} with corresponding patch positions shown in Fig.~\ref{fig:stateoftheart} for $P=4$.

        We consider a multi-patch sequence, where the excitation field is $T_R$-periodic within each patch. The time to move the patch from one position to the next is given by $T_{\shift}$ and we will consider the case where $T_{\shift}$ is a multiple of the drive-field excitation cycle $T_R$, i.e. $T_{\shift} = L_{\shift} T_R$ where $L_{\shift}$ is the number of drive-field excitation cycles required for patch movement. The data acquired during patch movement will not be considered for reconstruction. Thus, in order to reach a high duty cycle it is advantageous to measure for multiple drive-field excitation cycles at each patch. We consider a sequence where at each patch $L_P$ subsequent drive-field excitation cycles are measured such that the time per patch is given by $T_P = L_P T_R$. The duty cycle $D$ is given by
        \begin{equation}
            D = \frac{T_P}{T_P + T_{\shift}} = \frac{L_P}{L_P + L_{\shift}}.
        \end{equation}
        Consequently for reaching at least \SI{50}{\percent} duty cycle one should choose $L_P > L_\shift$. We note that $L_\shift$ has a scanner specific minimal value, which is determined by the possible slew-rate of the focus field.

        Getting back to Fig.~\ref{fig:multiPatchSequence} one can see that the sequence toggles between patch movement and measurement and proceeds until all patches are measured. We denote the repetition time of a full multi-patch cycle as
        \begin{equation}
            T_\text{frame} = P (T_P + T_\shift).
        \end{equation}
        In general, a multi-patch sequence will not only contain one frame but $F$ and in turn the total measurement time is
        \begin{equation}
            T_\text{meas} = F T_\text{frame} = L T_R,
        \end{equation}
        where $L=(L_P+L_\shift) P F$ is the total number of drive-field cycles, which we already used in the previous sections.

        Since $L_P$, $L_\shift$, $P$, and $F$ are all static parameters that do not change during the measurement it is possible to know precisely, in which patch the measurement sequence is for a certain time point $t \in [0,T_\text{meas}]$. To this end, we can define a mapping $\Pi:[0,T_\text{meas}] \rightarrow \{0,1,\dots,P\}$ that maps a time point to the patch number. As a special case we include patch number 0, which is returned in case that the patch is moving from one position to the next one. Taking this into account, the mapping can be defined as
        \begin{align}
            \Pi(t)=
            \begin{cases}
                \ceil{\frac{t \cmod {T_\frameN}P}{T_\frameN}} , 
                & t \cmod { \frac{T_\frameN}{P}}\leq T_R L_P\\ 
                0, & \text{else}
            \end{cases}\negthickspace
            \label{eq:mapping}
        \end{align}
        where $\ceil{\cdot}$ denotes the ceiling function. In a similar way we can define a function $\phi:[0,T_\text{meas}] \rightarrow \{1,\dots,F\}$ with
        \begin{equation} \label{eq:mappingFr}
            \phi(t)=\ceil{\frac{t}{T_\frameN}},
        \end{equation}
        which is mapping the time to the frame number.
        
    \subsection{Multi-Patch Data Combination}
        Now we switch back to the setting where not only the $F$ multi-patch frames are acquired, but also the tracer distribution is changing over time. For the reconstruction of a complete multi-patch image data from each patch are required. In order to generalize the data combination technique developed for single-patch imaging in section~\ref{spectralLeakage} we need take the patch substructure of a frame into account when constructing our virtual frame using the data snippets from section~\ref{sec:datasnippets}. 
        
        To this end we need to fill the virtual multi-patch frame $u_{t_s}^\text{virt}(t,\xi)$ with patch label $\xi$. The averaged virtual frame $\bar{u}_{t_s,h}^\virt(t,\xi)$ can be computed by equation~\eqref{eq:averagedAndSLCorr} with an adapted masking function
        \begin{align*}
            \tilde{\omega}_{t_s,h}^{\Delta t} :~&\IR\times\set{0,1,\dots,P} \longrightarrow [0,1],\\
            &(t,\xi)\mapsto 
            \begin{cases} \omega_{t_s,h}^{\Delta t}(t), & \xi = \Pi(t)\\ 0, & \text{else}\end{cases}.
        \end{align*}
        For successful reconstruction it is required that for all patches and all motion states, the entire drive-field cycle is captured after combining and weighting of the data, i.e. there exists no tuple $(t,\xi)$, $t\in[0,T_R]$ and $\xi \in \set{1,2,\dots,P}$, such that $\tilde{\omega}_{t_s,h}^{\Delta t}(t,\xi)=0$.

\section{Materials and Methods}
    \subsection{Experiments} \label{Sec:Experiments}
        \begin{figure}
            \centering
            \begin{tikzpicture}[node distance=2pt,every node/.style={inner sep=0pt}]
                \node[] (phantom) at (0,0) {\includegraphics[width=0.3\textwidth,trim=200 0 200 150,clip,scale=0.3]{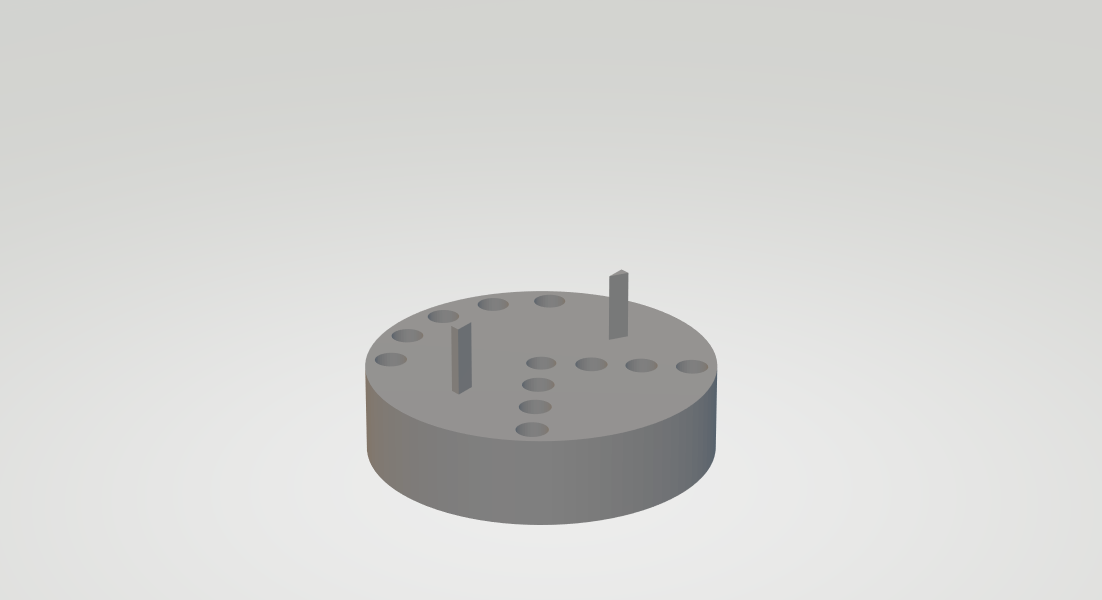}};
                \draw[|-|] (-1.7,-1.5) -- (1.6,-1.5) node[inner sep = 2pt,below, midway] {\SI{7}{\centi \meter}};
                \node[right = of phantom] (phantom2) {\includegraphics[width=0.175\textwidth]{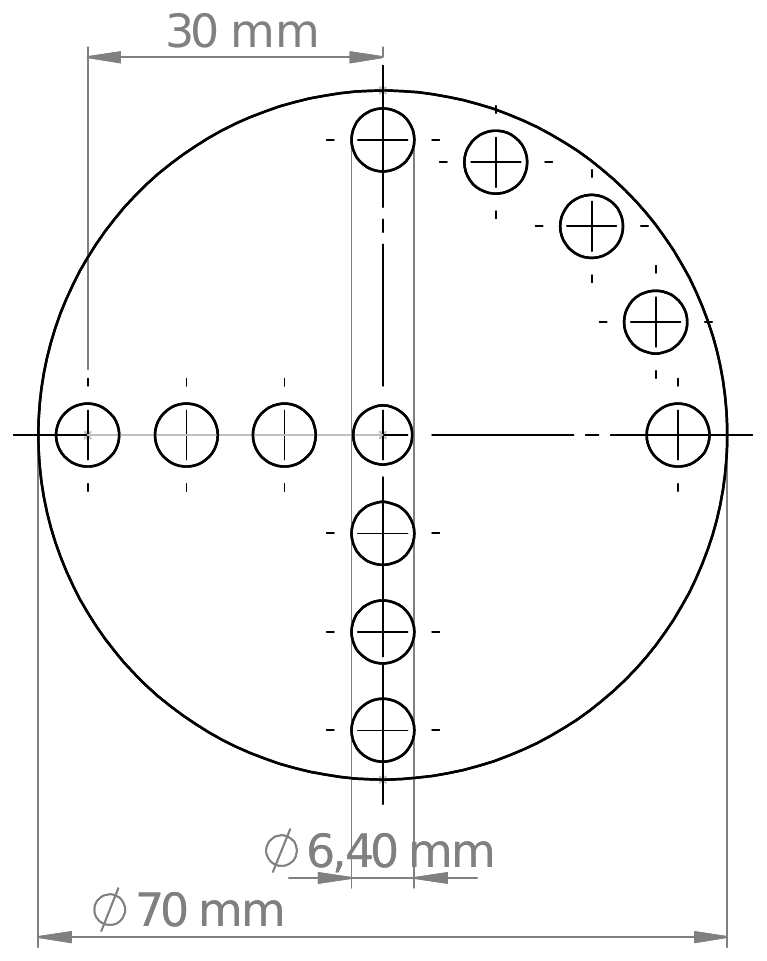}};
                \node[below = of phantom2,xshift=-2.8cm] () {\includegraphics[width=0.48\textwidth]{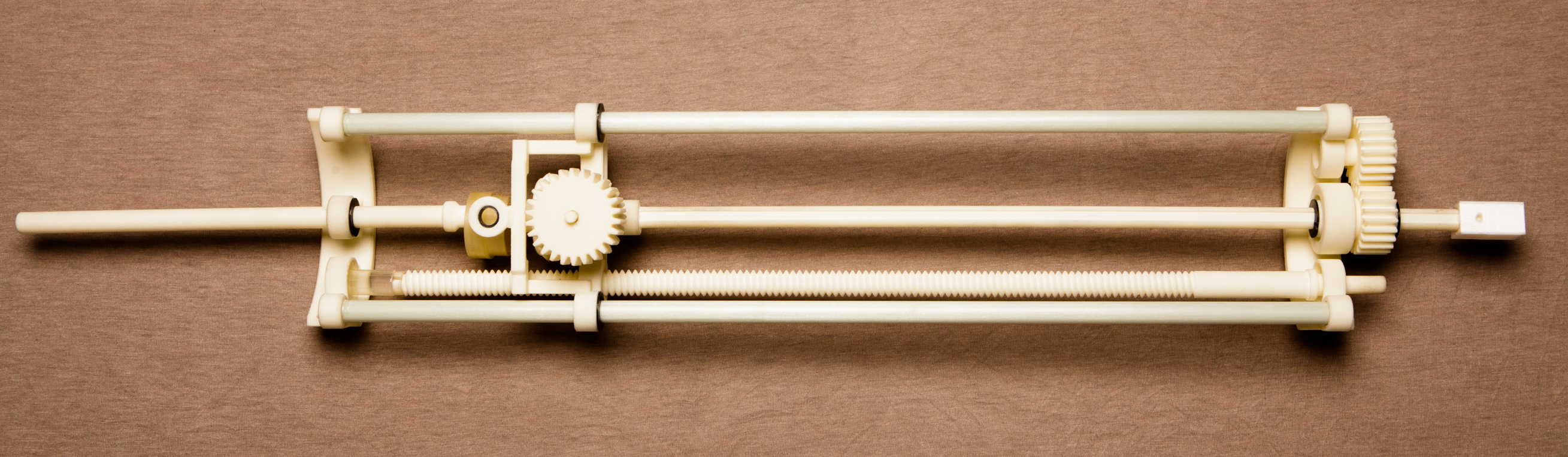}};
            \end{tikzpicture}
            \caption{Rotational phantom used in the experiments. The phantom (upper left and right) is attached to a phantom holder shown in the lower part of the figure. Eleven samples filled with \SI{10}{\micro \liter} of perimag are placed inside the holes of the phantom. By attaching a motor to the phantom holder, the phantom can be rotated. }
            \label{fig:phantom}
        \end{figure}
        Experiments were performed with a preclinical MPI scanner (Bruker, Ettlingen, Germany). We used a 3D-printed phantom holder with a diameter of \SI{7}{\centi \meter} shown in Fig.~\ref{fig:phantom} and placed eleven spherical shaped samples into the holes with the central hole left empty. Each sample had an inner diameter of about \SI{3}{\milli\meter} filled with \SI{10}{\micro \liter} perimag (micromod Partikeltechnologie GmbH, Rostock, Germany) with a concentration of \SI{50}{\milli \mol \of{Fe} \per \liter}. The phantom was placed horizontally inside the scanner on a phantom holder (Fig.~\ref{fig:phantom} bottom). A motor was attached to the phantom holder enforcing rotation in the horizontal plane inside the scanner bore of the phantom. 
        
        Multi-patch measurements were performed with the phantom in three different setups.
        \begin{itemize}
            \item With a static phantom, where the motor was turned off.
            \item With a slowly rotating phantom, where the average rotational frequency was \SI{0.814+-0.003}{\hertz} representing a case where the tracer distribution is quasi-static throughout a single drive-field cycle.
            \item With a fast rotating phantom, where the average rotational frequency was \SI{1.771+-0.003}{\hertz} representing a case where reconstructions of single excitation cycles would show motion artifacts.
        \end{itemize}
        Note that the frequencies were determined a posteriori from the raw measurement signal as described in section~\ref{sec:FrequencyDetermination}.
        
        All multi-patch measurements were performed with four patches, a gradient strength of \SI{-0.5}{\tesla \per \meter} in $x$- and $y$-direction, \SI{1}{\tesla \per \meter} in $z$-direction, and a drive-field amplitude of \SI{12}{\milli \tesla} in all three directions. The repetition time $T_R$ of a single drive field cycle was \SI{21.54}{\milli\second}. The four patches were shifted $\pm \SI{16}{\milli\meter}$ in $x$- and $y$-direction resulting in a FoV of \SI{80x80x24}{\milli \meter} as shown in Fig.~\ref{fig:stateoftheart}. At each patch position $L_P=\num{200}$ subsequent drive-field cycles were acquired, followed by seven drive-field cycles to change the focus fields to the next patch setting. We measured $2$ frames resulting in a total measurement time of $T_\meas=\SI{35.67}{\second}$ for each phantom. Thus, the slow and fast rotating phantoms cycle $29$ respectively $63$ time during the measurement. In addition to the phantom data we acquired $200$ background frames with an empty scanner bore and the motor turned off, which were averaged and subtracted from the phantom data~\cite{them2016sensitivity,knopp2019correction}. For the joint reconstruction, four system matrices were acquired with a delta sample of size \SI{2x2x1}{\milli \meter} on a grid of size \num{33x33x27} covering a volume of \SI{66x66x27}{\milli \meter} centered around the respective patch center shown in Fig.~\ref{fig:stateoftheart}.

    \subsection{Spectral Analysis}     \label{sec:FrequencyDetermination}
        The motor we used to rotate the phantom did not provide a method to accurately control or measure its frequency. Thus $\rho(t,n)$ and hence the window functions defined in \eqref{eq:window} are unknown at this point. Here, we describe how the frequency of motion can be determined by a spectral analysis of the raw MPI signal. 
        
        Experimentally, the continuous raw data signal $u(t)$ is sampled with a finite bandwidth for each receive channel. We used the signal of the $y$-channel for further analysis. To recover the motion frequency $\nu(\xi,f)$ for each patch $\xi\in\set{1,\dots,P}$ within each frame $f\in\set{1,\dots,F}$, we Fourier transform the time data acquired during each drive-field cycle. The resulting $K$ frequency components can be combined into a vector $\hat{\zb u}^{l,\xi,f} = \big( \hat{u}^{l,\xi,f}_k\big)_{k=1}^{K}$ for each drive-field cycle $l\in\set{1,\dots,L_P}$, patch and frame. For a fixed frequency component $k$, patch $\xi$, and frame $f$ the signal $\hat{u}^{l,\xi,f}_k$ can be interpreted as a time signal with $l$ being the time variable with a spacing of $T_R$ from which we generate a spectrum using the short-time Fourier transform. We then select the frequency component $k$ with the highest peak in the spectrum for determination of the motion frequency. We detect the fundamental $w=1$ and higher $w>1$ harmonic peaks of the motion frequency and use a parabolic interpolation technique with a Hann window~\cite{gasior2004} to determine their position $\nu_w(\xi,f)$. From there on we can calculate the motion frequency of the motor for any given patch $\xi$ and frame $f$ by
        \begin{equation}
            \nu(\xi,f) = \frac{\nu_w(\xi,f)}{w}
        \end{equation}
        with an uncertainty of
        \begin{equation}
            \Delta \nu = \frac{0.0526}{T_R L_P w}.
            \label{eq:freqerror}
        \end{equation}
        In our scenario the fourth harmonic was used which leads to an uncertainty of \SI{0.012}{\hertz}.
        
        The motion period $T_\mot(\xi,f)=\frac{1}{\nu(\xi,f)}$ is then used to recursively recover the function $\rho(t_s,n+1)$ by
        \begin{equation*}
            \rho(t_s,n+1)=\rho(t_s,n)+T_\mot(\Pi(\rho(t_s,n)),\phi(\rho(t_s,n))),
        \end{equation*}
        with $\Pi$ defined in \eqref{eq:mapping} and $\phi$ defined in \eqref{eq:mappingFr}.
        
    \subsection{Virtual Frames}
        Discretization is done by choosing the parameters $t_s$ and $\Delta t$. In our case, the motion states are chosen independently of the window width $\Delta t$.
        
        During one rotation of the phantom multiple drive-field cycles $M=\floor{\frac{T_\mot}{T_R}}$ are performed, where $T_\mot$ is the mean motion period averaged over all patches and frames and $\floor{\cdot}$ denotes the floor function. In our scenario we choose to visualize the phantom at each drive-field cycle and hence
        \begin{align*}
            t_s \in \set{ m T_R : m =0,\dots, M-1 }
        \end{align*}
        with $M$ equal to \num{29} and \num{63} for slow and fast rotation, respectively. The window widths $\Delta t$ are chosen from $0.6T_R$ to $3.0T_R$ in increments of $0.3T_R$. Virtual frames are computed for all possible combinations of $t_s$, $\Delta t$ and the Hann respectively the rectangular window function. The only exception being $\Delta t = 0.6T_R$ for the slowly rotating phantom as the virtual frame is undefined in this case.

    \subsection{Image Reconstruction}     \label{sec:Reconstruction}
        Once the virtual frames are computed for a specific state of motion, an image reconstruction technique is required to determine the corresponding tracer distribution. Here, we use the joint multi-patch reconstruction approach~\cite{Knopp2015PhysMedBiol, szwargulski2018efficient}, which makes use of a regularized version of the iterative Kaczmarz algorithm~\cite{Knopp2010PhysMedBio}. Moreover, for the reconstruction we use only frequencies above \SI{80}{\kilo\hertz} with an signal-to-noise ratio (SNR) greater than $2$ to account for background noise~\cite{them2016sensitivity}.
        
        Reconstruction parameters are chosen manually to minimize the blurring of the samples. For reference, the measurement data of the static and the fast rotating phantom are fully averaged and reconstructed with a relative Tikhonov regularization parameter $\lambda = 0.001$ and $5$ iterations. The virtual frames obtained from the measurement of the dynamic phantom are reconstructed with a relative Tikhonov regularization parameter of $\lambda = 0.01$ and $2$ iterations.
        
        For visualization and analysis we convert the reconstructed 3D MPI tomograms into 2D images by maximum intensity projections in $z$-direction. Additionally, we generate two movies for each rotating phantom from all motion states. For the slower rotating phantom we use $\Delta t=1.5T_R$ and for the fast rotating phantom $\Delta t=0.9T_R$ and the Hann window. The first movie shows the rotation in real time with $46$ frames per second, while the second is decelerated to $4$ frames per second to highlight the phantom displacement from frame to frame.

    \subsection{Image SNR and Sample Width}
        Motion artifacts appear most prominent for the samples on the diagonal of the rotating phantom with the highest distance from the rotational center. Therefore, motion artifacts are quantified by the full width at half maximum (FWHM) of these samples. The FWHM is calculated for all $t_s$ where the diagonals of the phantom are approximately aligned horizontally respectively vertically. To this end, the image of the static phantom is slightly rotated before the FWHM is obtained. Image noise is quantified by the SNR. A circular mask with a radius of $17$~pixels is used to cover the central region. This is the region where the phantom is located. The maximal pixel value within this region defines our signal, whereas the noise is given by the standard deviation of all pixel values outside the mask. The SNR is calculated for all $t_s$.

\section{Results} \label{Sec:Results}
    \begin{figure}
        \centering
        \input{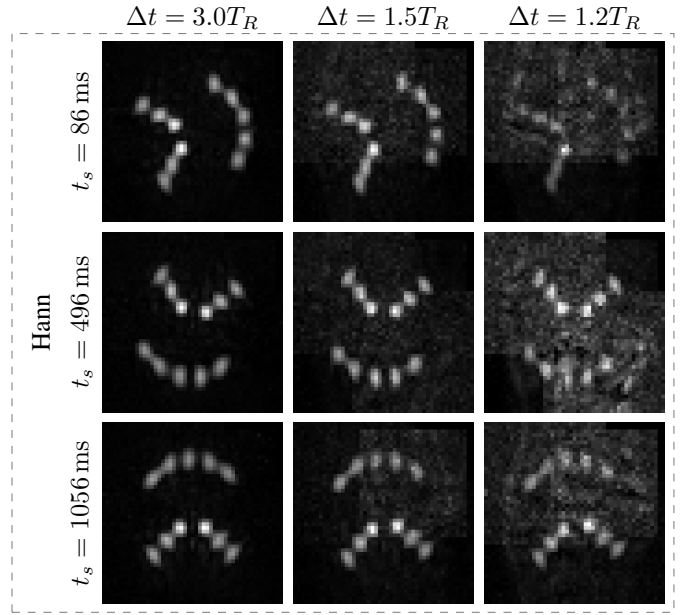}
        \caption{Motion-compensated reconstructions of the slowly rotating phantom are depicted. In all cases a Hann window is used for spectral leakage correction. Images within the same row show the same motion state labeled by a specific time point $t_s$ in the first motion cycle which has a duration of \SI{1229}{\milli\second}. The columns differ in the width of the data snippets $\Delta t$ used for reconstruction. The widths of $3.0T_R$, $1.5T_R$, and $1.2T_R$ are approximately equal to \SIlist{65;45;26}{\milli\second}, respectively. These are \SIlist{5.3;3.7;2.1}{\percent} of the duration of the motion cycle. One observes that the suppression of motion artifacts improves with decreasing $\Delta t$, while the image noise increases.}
        \label{fig:results1}
    \end{figure}
    
    \begin{figure}
        \centering
        \input{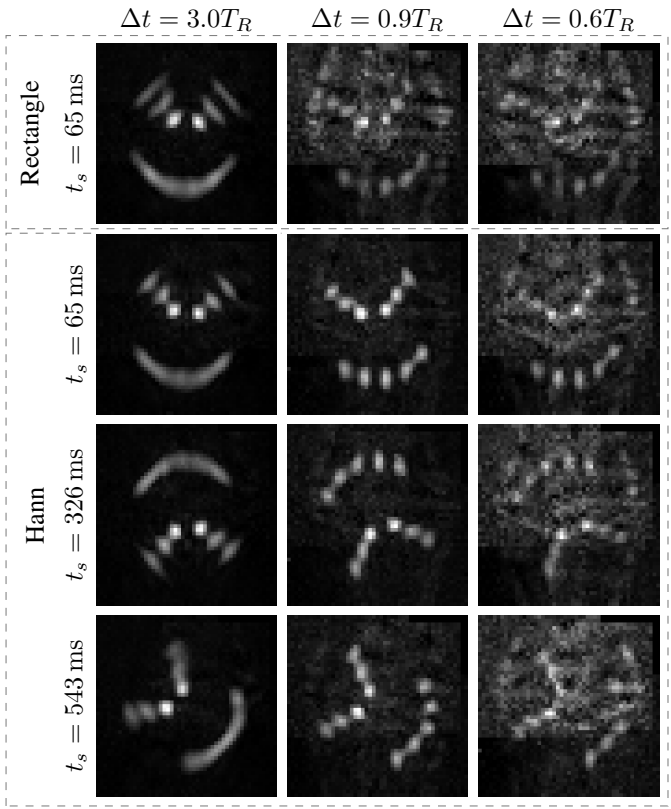}
        \caption{Motion-compensated reconstructions of the fast rotating phantom are depicted. The three images at the top use a rectangular window which is equivalent to no spectral leakage correction, whereas a Hann window is used in all other cases. Images within the same row show the same motion state labeled by a specific time point $t_s$ in the first motion cycle which has a duration of \SI{565}{\milli\second}. The columns differ in the width of the data snippets $\Delta t$ used for reconstruction. The widths of $3.0T_R$, $0.9T_R$, and $0.6T_R$ are approximately equal to \SIlist{65;19;13}{\milli\second} which is \SIlist{11.5;3.4;2.3}{\percent} of the duration of the motion cycle. Just as with the slowly rotating phantom shown in Fig.~\ref{fig:results1} one observes that the suppression of motion artifacts improves with decreasing $\Delta t$, while the image noise increases. Additionally, one observes that image noise decreases with the use of spectral leakage correction (Hann) as is best shown for $\Delta t=0.9T_R$. Artifact suppression on the other hand is inferior without spectral leakage correction (Rectangle) as best seen for $\Delta t=3.0T_R$, where structures on the diagonal are visibly wider compared to the images using the Hann window function.}
        \label{fig:results2}
    \end{figure}
    
    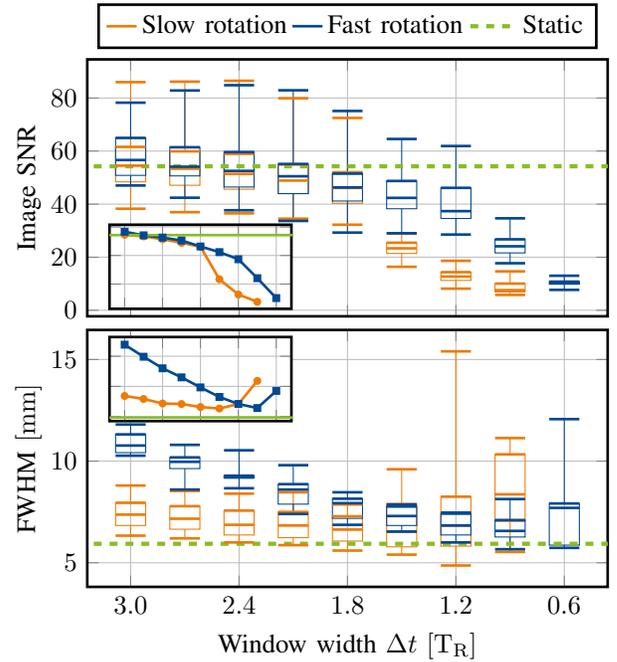
\begin{figure}
        \centering
        \hspace{-1.25cm}
        \begin{tikzpicture}

\pgfplotsset{every axis/.append style={line width=1pt},
             ylabel style={at={(-0.1,0.5)},anchor=base},}

\pgfplotsset{
    box plot/.style={
        /pgfplots/.cd,
        black,
        only marks,
        mark=-,
        mark size=\pgfkeysvalueof{/pgfplots/box plot width},
        /pgfplots/error bars/y dir=plus,
        /pgfplots/error bars/y explicit,
        /pgfplots/table/x index=\pgfkeysvalueof{/pgfplots/box plot x index},
    },
    box plot box/.style={
        /pgfplots/error bars/draw error bar/.code 2 args={%
            \draw  ##1 -- ++(\pgfkeysvalueof{/pgfplots/box plot width},0pt) |- ##2 -- ++(-\pgfkeysvalueof{/pgfplots/box plot width},0pt) |- ##1 -- cycle;
        },
        /pgfplots/table/.cd,
        y index=\pgfkeysvalueof{/pgfplots/box plot box top index},
        y error expr={
            \thisrowno{\pgfkeysvalueof{/pgfplots/box plot box bottom index}}
            - \thisrowno{\pgfkeysvalueof{/pgfplots/box plot box top index}}
        },
        /pgfplots/box plot
    },
    box plot top whisker/.style={
        /pgfplots/error bars/draw error bar/.code 2 args={%
            \pgfkeysgetvalue{/pgfplots/error bars/error mark}%
            {\pgfplotserrorbarsmark}%
            \pgfkeysgetvalue{/pgfplots/error bars/error mark options}%
            {\pgfplotserrorbarsmarkopts}%
            \path ##1 -- ##2;
        },
        /pgfplots/table/.cd,
        y index=\pgfkeysvalueof{/pgfplots/box plot whisker top index},
        y error expr={
            \thisrowno{\pgfkeysvalueof{/pgfplots/box plot box top index}}
            - \thisrowno{\pgfkeysvalueof{/pgfplots/box plot whisker top index}}
        },
        /pgfplots/box plot
    },
    box plot bottom whisker/.style={
        /pgfplots/error bars/draw error bar/.code 2 args={%
            \pgfkeysgetvalue{/pgfplots/error bars/error mark}%
            {\pgfplotserrorbarsmark}%
            \pgfkeysgetvalue{/pgfplots/error bars/error mark options}%
            {\pgfplotserrorbarsmarkopts}%
            \path ##1 -- ##2;
        },
        /pgfplots/table/.cd,
        y index=\pgfkeysvalueof{/pgfplots/box plot whisker bottom index},
        y error expr={
            \thisrowno{\pgfkeysvalueof{/pgfplots/box plot box bottom index}}
            - \thisrowno{\pgfkeysvalueof{/pgfplots/box plot whisker bottom index}}
        },
        /pgfplots/box plot
    },
    box plot median/.style={
        /pgfplots/box plot,
        /pgfplots/table/y index=\pgfkeysvalueof{/pgfplots/box plot median index}
    },
    box plot width/.initial=1em,
    box plot x index/.initial=0,
    box plot median index/.initial=1,
    box plot box top index/.initial=2,
    box plot box bottom index/.initial=3,
    box plot whisker top index/.initial=4,
    box plot whisker bottom index/.initial=5,
}

\newcommand{\boxplot}[2][]{
    \addplot [box plot median,#1] table [col sep=comma] {#2};
    \addplot [forget plot, box plot box,#1] table [col sep=comma] {#2};
    \addplot [forget plot, box plot top whisker,#1] table [col sep=comma] {#2};
    \addplot [forget plot, box plot bottom whisker,#1] table [col sep=comma] {#2};
}

\begin{axis} [ 
    grid=both, 
    width=8.5cm,
    height=5cm,
    box plot width=2mm,
    ylabel = Image SNR,
    x dir=reverse,
    xtick={0.6,1.2,...,3.0},
    xticklabels={},
    legend style={at={(0.5, 1.04)},anchor = south},
    legend columns=3,
    ]
    \boxplot [ukesec2,
        box plot whisker bottom index=1,
        box plot whisker top index=5,
        box plot box bottom index=2,
        box plot box top index=4,
        box plot median index=3
    ] {data/gdani8a.txt};
    \addlegendentry{Slow rotation};
    
    \boxplot [ibidark,
        box plot whisker bottom index=1,
        box plot whisker top index=5,
        box plot box bottom index=2,
        box plot box top index=4,
        box plot median index=3
    ] {data/gdani8b.txt};
    \addlegendentry{Fast rotation};
    
    \addplot[ukesec4,update limits=false, on layer=background,dashed, ultra thick] coordinates {(0,54.26299766570726) (4,54.26299766570726)}; 
    \addlegendentry{Static};
\end{axis}

\begin{axis} [
    yshift=-3.6cm,
    width=8.5cm,
    height=5cm,
    grid=both, 
    box plot width=2mm,
    ylabel = FWHM,
    y unit prefix = m, y unit = m,
    xlabel = Window width $\Delta t$,
    x unit = T_R,
    x dir=reverse,
    xtick={0.6,1.2,...,3.0},
    x tick label style={
            /pgf/number format/.cd,
                fixed,
                fixed zerofill,
                precision=1,
            /tikz/.cd
        },
    ]
    
    \boxplot [ukesec2,
        box plot whisker bottom index=1,
        box plot whisker top index=5,
        box plot box bottom index=2,
        box plot box top index=4,
        box plot median index=3
    ] {data/gdani8c.txt};
    
    \boxplot [ibidark,
        box plot whisker bottom index=1,
        box plot whisker top index=5,
        box plot box bottom index=2,
        box plot box top index=4,
        box plot median index=3
    ] {data/gdani8d.txt};
    
    \addplot[ukesec4,update limits=false, on layer=background,dashed,ultra thick] coordinates {(0,5.933333333333417) (4,5.933333333333417)}; 
\end{axis}

\node at (-0.75,0.1) {\begin{axis}[
    grid=both,
    width=4cm,
    height=2.7cm,
    yticklabels = {},
    x dir=reverse,
    xtick={0.6,1.2,...,3.0},
    xticklabels={},
    y label style={at={(-0.1,0.5)}},
    axis background/.style={fill=white},
    mark options={scale=0.5},
    ]
    
    \addplot [ukesec2, mark=*] table [x index = 0  ,y index = 3,col sep=comma] {data/gdani8a.txt};
    \addplot [ibidark, mark=square*] table [x index= 0 ,y index = 3,col sep=comma] {data/gdani8b.txt};
    \addplot [ukesec4,sharp plot,update limits=false] coordinates {(0,54.26299766570726) (3.5,54.26299766570726)};
\end{axis}};

\node[anchor=north] at (-0.75,-1.27) {\begin{axis}[
    grid=both,
    width=4cm,
    height=2.7cm,
    ymin = 5.7,
    yticklabels = {},
    x dir=reverse,
    xtick={0.6,1.2,...,3.0},
    xticklabels={},
    y label style={at={(-0.1,0.5)}},
    axis background/.style={fill=white},
    mark options={scale=0.5},
    ]
    
    \addplot [ukesec2, mark=*] table [x index = 0  ,y index = 3,col sep=comma] {data/gdani8c.txt};
    \addplot [ibidark,mark=square*] table [x index= 0 ,y index = 3,col sep=comma] {data/gdani8d.txt};
    \addplot [ukesec4,sharp plot,update limits=false] coordinates {(0,5.933333333333417) (3.5,5.933333333333417)};
\end{axis}};

\end{tikzpicture}
        \caption{Signal-to-noise ratio (top) as a measure of image noise and full width at half maximum (bottom) of the samples on the radial structure with the highest distance from the rotational center as measure for the sample width are shown depending on the window widths $\Delta t$. For the slow and fast rotating phantom SNR and FWHM are computed for different points in time which are collected in orange and blue box plots, respectively. For the static phantom the median SNR and FWHM values are shown as a dashed green line. Additionally, the median values are shown in the smaller extra plots. The median image SNR for the rotating phantoms starts at about the median SNR of the static phantom for window widths above $2.4T_R$ and drops off quickly below $\Delta t=1.8T_R$ for the slow phantom and $\Delta t=1.2T_R$ for the fast phantom. For both rotations the median FWHM declines until about the FWHM of the static phantom as one shortens the window width $\Delta t$ until shortly after the drop-off point of the SNR from where on it increases.}
        \label{fig:snr}
    \end{figure}
    
    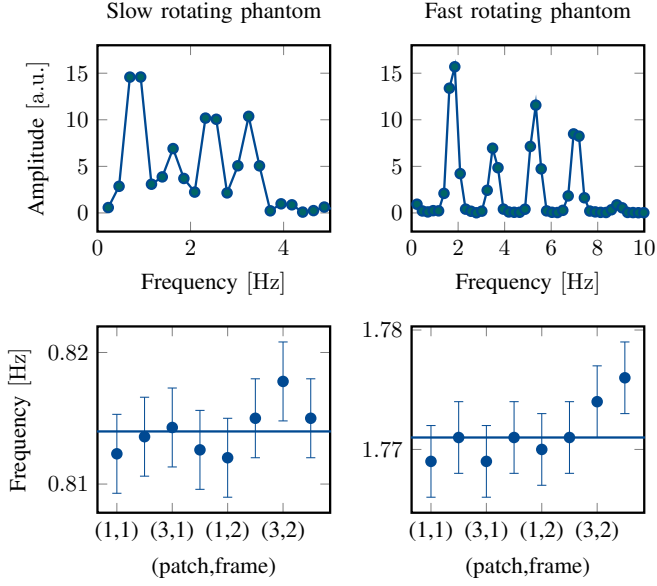
\begin{figure}
        \centering
        \begin{tikzpicture}[scale=0.87]

\pgfplotsset{every axis/.append style={line width=1pt},
                                       cycle list name = exotic,
                                       width=0.58*\columnwidth
                                      }
                                      
\begin{axis}[
    xshift=4.8cm,yshift=0cm,
    use units,
    xlabel = Frequency,
    x unit = Hz,
    scaled ticks=false,
    scaled y ticks=manual:{}{\pgfmathparse{#1/(10^6)}},
    title = Fast rotating phantom,
    legend pos = south east,
    xmin=0,xmax=10,
    ymin=-2*10^6,ymax=18*10^6,]
    
    \addplot+ [ibidark] table [x index = 0  ,y index = 1] {data/gdani9a.txt};
\end{axis}

\begin{axis}[
    use units,
    ylabel = Amplitude,
    xlabel = Frequency,
    x unit = Hz,
    y unit = a.u.,
    scaled y ticks=manual:{}{\pgfmathparse{#1/(10^6)}},
    title = Slow rotating phantom,
    legend pos = south east,
    xmin=0,xmax=5,
    ymin=-2*10^6,ymax=18*10^6,]
    
    \addplot+ [ibidark] table [x index = 0  ,y index = 1] {data/gdani9b.txt};
\end{axis}

\begin{axis}[
    xshift=0cm,yshift=-4.3cm,
    use units,
    xlabel = {(patch,frame)},
    ylabel = Frequency,
    y unit = Hz,
    xtick={1,3,5,7},
    xticklabels={(1,1),(3,1),(1,2),(3,2)},
    ytick distance = 0.01,
    ]
    
    \addplot[ibidark,only marks, error bars/.cd, y dir=both, y explicit] table[x index = 0  ,y index = 1, y error index = 2] {data/gdani9c.txt};
    \addplot[ibidark, no marks, update limits=false] coordinates {(0,0.814) (9,0.814)};
\end{axis}

\begin{axis}[
    xshift=4.8cm,yshift=-4.3cm,
    use units,
    xlabel = {(patch,frame)},
    xtick={1,3,5,7},
    xticklabels={(1,1),(3,1),(1,2),(3,2)},
    ytick distance = 0.01,
    ]
    
    \addplot [ibidark,only marks, error bars/.cd, y dir=both, y explicit] table[x index = 0, y index = 1, y error index = 2] {data/gdani9d.txt};
    \addplot [ibidark, no marks,update limits=false] coordinates {(0,1.771) (9,1.771)};
\end{axis}
\end{tikzpicture}
        \caption{Here, the determination of the motion frequencies from the raw data is summarized for the slowly rotating phantom (left) and the fast rotating phantom (right). Exemplary, the spectra for the first frame and patch are shown (top). In both cases, up to four harmonics of the motion frequency show a prominent peak. The peak of the fourth harmonics is used to determine the motion frequency (blue marker) for each patch and frame (bottom), as it provides the best accuracy (error bar) of \SI{0.003}{\hertz}. The mean frequencies are indicated by a blue line at \SI{0.814}{\hertz} for the slow rotation and \SI{1.771}{\hertz} for the fast rotation.}
        \label{fig:spectrum}
    \end{figure}
    
    \begin{figure}
        \centering
        \input{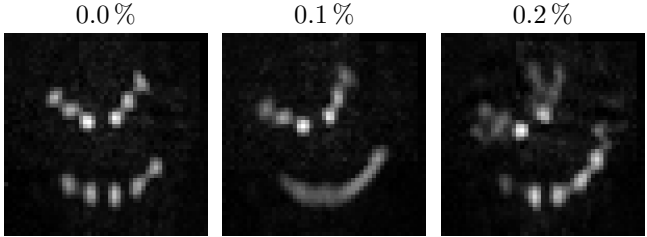}
        \caption{Reconstructed images of the fast rotating phantom at $t_s=\SI{65}{\milli \second}$ with $\Delta t =1.0T_R$ are shown, where the motion frequency of the phantom is deliberately overestimated by \SIlist{0.1;0.2}{\percent}. The reconstruction using the reference frequency is shown on the left hand side. In our example even small overestimation and underestimation (not shown) lead to smearing (center) and even ghosting artifacts highlighting the importance of an accurate spectral analysis.}
        \label{fig:harmonic}
    \end{figure}
   
    As reference, we consider a state-of-the-art reconstruction of the static and fast rotating phantom shown in Fig.~\ref{fig:stateoftheart}. While the phantoms structure is perfectly visible in the static case it is completely lost in the dynamic one. Motion-compensated reconstructions of the slowly and fast rotating phantom are shown in Fig.~\ref{fig:results1} and~Fig.~\ref{fig:results2}, respectively. Our method has two free parameters which influence how strong motion artifacts are suppressed. These are the window function $h$ and the window width $\Delta t$ which is equal to the length of our data snippets. Visual inspection of the motion-compensated reconstructions shows that the suppression of motion artifacts improves with decreasing window width, whereas image noise increases. The impact of the spectral leakage correction is illustrated in~Fig.~\ref{fig:results2}, where reconstructions without (rectangular window) and with (Hann window) correction are shown. We observe that for a fixed window width more image noise and stronger motion artifacts are present without spectral leakage correction. Furthermore, we notice in the movies that the image noise depends on the patches and is not homogeneously distributed over the entire FoV.

    Using the Hann window we perform a quantitative analysis of artifact suppression measured by the FWHM and image noise measured by the image SNR. The results are summarized in Fig.~\ref{fig:snr}. For both the fast and the slowly rotating phantom, the median SNR values reach up to the median SNR value of the static phantom of $54$ for large window widths of $2.4T_R$ and higher. They then decline slowly as the window width decreases until they reach a drop-off point from where they fall off quickly agreeing well with our visual observations. The slowly rotating phantom reaches this drop-off point at $\Delta t = 1.8$ and the fast rotating phantom at $\Delta t = 1.2$. The FWHM of the static phantom is \SI{6.0}{\milli\meter} and the FWHM of the rotating phantoms starts at \SIlist{10.8;7.4}{\milli\meter} for $\Delta t = 3.0\; T_R$ for fast and slow rotation, respectively. The FWHM then decreases with decreasing window width and reaches an optimum of \SI{6.6}{\milli\meter} at $\Delta t = 0.9\; T_R$ for the fast rotation and \SI{6.5}{\milli\meter} at $\Delta t = 1.5\; T_R$ for the slow rotation from which on it increases again.
    
    The determination of the motion frequencies from the raw data signal is summarized in Fig.~\ref{fig:spectrum}. For both the fast and the slowly rotating phantom the fundamental oscillation and a series of higher harmonics are visible in the spectrum. Up to the fourth harmonic the peaks in the spectrum are quite prominent and are well suited for the parabolic interpolation technique. Using the fourth harmonic the rotational frequencies are determined. The methodological error is \SI{0.003}{\hertz} as given by equation~\eqref{eq:freqerror}. Most frequencies lie within the uncertainty region around the average frequencies of \SI{0.814+-0.003}{\hertz} and \SI{1.771+-0.003}{\hertz} for the slow and fast rotation, respectively. The average motion period is thus given by \SI{1.229+-0.005}{\second} for the slow rotation and \SI{0.564+-0.001}{\second} for the fast one. 
    
    For a successful motion compensation it is crucial to accurately determine the motion frequency. In order to illustrate this point we have deliberately underestimated and overestimated the frequency of the fast rotation by \SIlist{0.1;0.2}{\percent} and applied our generalized motion compensation method. As shown in Fig.~\ref{fig:harmonic} for the overestimated case motion artifacts re-emerge compared to the reconstruction assuming the correct frequency.

\section{Discussion} \label{Sec:Discussion}
    This work proposes a raw data processing technique to reduce motion artifacts when imaging large tracer distributions experiencing periodic dynamics using MPI multi-patch sequences. Our technique operates on the raw data only, such that the virtual frames can be reconstructed with any reconstruction method. Potentially, even time-signal based gridding techniques as recently developed in~\cite{ozaslan2019fully} can be applied. With this approach, we were able to image a structured phantom at different states of a rotary motion in an almost motion artifact-free manner and recover its complete motion using the full multi-patch dataset. Our method thus provides a significant improvement over state-of-the-art techniques, which lead to a complete loss of the phantoms structure in our case.
    
    While our method provides a significant improvement, it also has its limitations. At first, we introduced the assumption that the tracer distribution needs to be temporally recurring, so the same state of motion reoccurs multiple times during the multi-patch sequence. This assumption is violated e.g. in dynamic bolus experiments, where the bolus is either taken up by the liver too fast or only a single bolus pass is of interest. In case of uptake, it should be possible to extend our method by incorporating a suitable model for the signal drop. In the latter case, one may circumvent motion artifacts in the first place by using specific imaging sequences optimized towards high temporal resolution~\cite{vogel2020}, which currently comes at the cost of restricting the FoV to 2D. The second assumption is that the virtual frame in equation~\eqref{eq:averagedAndSLCorr} needs to be defined, which limits the minimal measurement length depending on the window width, dynamic of the tracer distribution, and multi-patch sequence. Unfortunately, this minimal measurement length has no simple closed form expression or upper bound. However, in a best case single-patch scenario we can cover a virtual frame by a number of data snippets equal to the up rounded ratio between $T_R$ and effective length of the data snippets. Using a continuous measurement with \SI{100}{\percent} duty cycle we need to measure for at least that number of repetitions of our motion. For a multi-patch scenario, we need to perform these measurements at least once for each patch further increasing the minimal measurement duration. Exemplary, we need at least \num{229} and \num{105} measured drive-field cycles per patch for the slow and fast rotating sample with a Hann window with window length of $\Delta t = 0.6 T_R$ in our measurement scenario, which is well below the total number of \num{400} cycles available per patch. For the fast rotating sample the virtual frame is well defined. However, it is undefined for the slow rotating sample, indicating that the best case scenario provides only a rough estimate of the minimal measurement length. In detail also the specifics of the multi-patch sequence, i.e. whether the multi-patch sequence rapidly switches between patches or measures for some time at each patch position, do influence the minimal measurement length. Hence, it is possible that specific sequences could outperform others with respect to minimizing the measurement time. In the present work we did not focus on this particular question but used the multi-patch sequence with highest possible duty cycle, since the scanner hardware used in our experiment has a minimum time of seven drive-field cycles to change focus which drastically reduces duty cycle for rapidly shifting sequences.
    
    The proposed method introduces two parameters: the window width $\Delta t$ and the window function $h$. The window function allows to tune the spectral leakage correction. We have investigated the impact of the latter by a qualitative analysis. We found that both motion artifacts and noise are less prominent with spectral leakage correction. Furthermore, we quantitatively investigated the impact of window width $\Delta t$ on motion artifact suppression and image noise. For large window widths we observed as little image noise as in the reference reconstruction of the static phantom, while the noise increased for smaller widths. Artifact suppression on the other hand first improved with decreasing window width until the median sample width of the dynamic structured phantom is about \SI{10}{\percent} above the width of the static one and the corresponding image noise has dropped roughly by a factor of two. From there on the median width increases and the corresponding distribution of sample widths widens, which is caused by a significant increase of image noise. In summary, the window width impacts both artifact suppression and image noise, however with opposing trends. With the significant impact of window width and window function on artifact suppression and image noise the question remains how to choose these parameters. As for the window functions, the Hann window clearly outperformed the rectangular window, which highlights the significance of spectral leakage correction on the overall performance of our method. It provided very good results in our scenario but there are likely window functions, which  perform better. As for the window width, we already discussed the related trade-off between artifact suppression and image noise, which is why its choice is subjective and dependent on the specific imaging scenario. Moreover, we need to take into consideration that the total scan duration of the multi-patch sequence and the specific dynamics of the tracer distribution limits the minimum value of the window width since the virtual frame needs to be defined for all $t_s$.
    
    In our specific imaging scenario one drive-field cycle takes \SI{21.54}{\milli\second} during which samples on the outer arc of the phantom cover a distance of approximately $1.7$ and $3.6$~pixels in image space for slow and fast rotation, respectively. If we want to restrict the displacement below one pixel we have to limit the effective length of the data snippets to $0.6\,T_R$ respectively $0.3\,T_R$. This can be accomplished by choosing the window width $\Delta t$ such that the FWHM of the chosen window function equals this effective length. I.e. $\Delta t$ equals $1.2\,T_R$ respectively $0.6\,T_R$ for the Hann window. Since the image noise is already quite significant and the sample width is sub-optimal for these window widths, the values minimizing the sample width would be a better choice since the image noise is still not that prominent in this case.

    The recurrence time of the motion states was obtained from the raw MPI signal by a simple spectral analysis without any additional measurements. In our imaging scenario the motion frequency was determined for each patch individually to account for frequency variations in our experimental setup, which are of the same magnitude as the methodological error of the spectral analysis. We analyzed our methods sensitivity with respect to systematic underestimations or overestimations of the motion frequency and found that already quite small relative deviations of \SIlist{0.1;0.2}{\percent} lead to a reappearance of motion artifacts. These deviations are already smaller than the relative uncertainties of \SI{0.2}{\percent} respectively \SI{0.4}{\percent} with which the motion frequency was determined. In cases where the distribution changes with a larger number of faster varying oscillatory modes or focus field shifts occur more frequently more sophisticated spectral analysis methods should be considered~\cite{clemson2016,iatsenko2015,sheppard2011}.
    
    In conclusion, our method allows to reduce artifacts from tracer distributions with periodic dynamics without making further assumptions on the dynamics and without modifications to the multi-patch measurement sequence. Because no image registration is involved and the spectral analysis is performed on the raw data, the numerical costs of our method are comparable to a standard multi-patch reconstruction. The proposed method is of high importance for those \textit{in-vivo} scenarios, where respiration and heart beat are common sources for motion artifacts. This is especially true when considering human applications, where a multi-patch imaging sequence will be required due to the increasing size of the measurement field.

\appendix
    \subsection{Implementation} \label{Sec:Implementation}        
        All algorithms are implemented in the programming language Julia (version 1.2)~\cite{bezanson2017julia} and integrated into the MPI software package \texttt{MPIReco.jl} (version 0.1.1)~\cite{knopp2019mpireco}. An example script that can be adapted to reproduce all reconstruction results shown in this paper can be found in the directory \texttt{examples/MotionCompensation} of the software package. The example can also be accessed through the web link \url{https://github.com/MagneticParticleImaging/MPIReco.jl/tree/master/examples/MotionCompensation}. 
        
        All measured data is stored in the MDF data format (version~2.1)~\cite{MDF2018} and can be accessed using the Julia package \texttt{MPIFiles.jl} (version 0.7.1)~\cite{knopp2019mpifiles}. The data is integrated into the OpenMPIData platform~\cite{OpenMPIData} under a Creative Common license. The aforementioned example script will automatically download the data and perform a reconstruction. For convenience, the system matrices are provided in two ways. First, in an uncompressed form with a frequency selection (SNR threshold of $2$) applied such that the file size is reduced from about \SI{34.1}{\giga\byte} to \SI{2.4}{\giga\byte}. Since we use the same threshold for reconstruction, all MPI tomograms shown in this manuscript can be reproduced. In addition, we provide compressed system matrices with a size of about \SI{245}{\mega\byte}. Since the compression is lossy, the images will appear slightly different than the ones in this publication. Within the example reconstruction script one can choose, which system matrices are downloaded and used.

\bibliographystyle{IEEEtran}

\bibliography{ref}

\end{document}